\documentclass[onecolumn,showpacs,nopreprintnumbers,aps,prd, letterpaper,groupeaddress,nofootinbib,tightenlines,floats,floatfix,showkeys]{revtex4-2}
\usepackage{graphicx}
\usepackage{hyperref}
\usepackage[english]{babel}
\usepackage{amsmath,amssymb}
\usepackage{slashed}
\usepackage{appendix}
\usepackage{orcidlink}
\usepackage[caption=false]{subfig}
 
\begin{document}

\title{On the role of magnetic fields into the dynamics and gravitational wave emission of binary neutron stars}

\author{Mariana Lira\orcidlink{00-0002-1825-205X}}
%\email{lira@ciencias.unam.mx} 
\altaffiliation{Instituto de Ciencias Nucleares, Universidad Nacional Aut\'onoma de M\'exico,
Circuito Exterior C.U., A.P. 70-543, M\'exico D.F. 04510, M\'exico}
\author{Juan Carlos Degollado\orcidlink{0000-0002-8603-5209}}
%\email{jcdegollado@ciencias.unam.mx}
\altaffiliation{Instituto de Ciencias F\'isicas, Universidad Nacional Aut\'onoma de M\'exico,
Apdo. Postal 48-3, 62251, Cuernavaca, Morelos, M\'exico}

\author{Claudia Moreno\orcidlink{0000-0002-0496-032X}}
%\email{jcdegollado@ciencias.unam.mx}
\altaffiliation{Departamento de F\'isica,
Centro Universitario de Ciencias Exactas e Ingenier\'ia, Universidad de Guadalajara\\
Av. Revoluci\'on 1500, Colonia Ol\'impica C.P. 44430, Guadalajara, Jalisco, M\'exico}

\author{Darío Núñez\orcidlink{0000-0003-0295-0053}}
\email{nunez@nucleares.unam.mx}
\altaffiliation{Instituto de Ciencias Nucleares, UNAM}

\date{\today}

\begin{abstract}
Modelling as a dipole the magnetic interaction of a binary system of neutron stars, we are able to include the magnetic effects in the Newtonian and in the inspiral dynamics of the system 
using an equivalent one-body description. Furthermore, in the inspiral stage  
we determine the role of the magnetic interaction in the waveforms generated by the system and obtain explicit formulas for the decrease in the separation of the stars, the time to reach a minimal radius, the gravitational luminosity and the change of gravitational wave frequency, all this within the quadrupole approximation. For the magnitude of the magnetic field that is consider to exist in these binaries $\sim 10^{16} {\rm G}$ 
we are able to show that its effect on the observable quantities is of the order of the 2PN correction, already close to the detection range of the gravitational waves observatories.
%}
We also discuss cases in which the magnetic field could have a more significant influence.

\end{abstract}
\maketitle

\section{INTRODUCTION}

 Gravitational Wave (GW) astronomy is becoming one of the most promising fields in astrophysics since the first detection of a binary black hole system GW150914 \cite{LIGOScientific:2016aoc} by the LIGO and Virgo Collaboration (LVC) \cite{LIGOScientific:2007fwp,LIGOScientific:2014pky}. 
Furthermore, other detections such as GW170817 \cite{LIGOScientific:2017vwq,LIGOScientific:2017ync} and GW190425 \cite{LIGOScientific:2020aai},
are consistent with the GWs emitted during the last orbits of a Binary Neutron Stars (BNS) coalescence. 

Currently, BNS are among the main sources of GWs studied by the new born multi-messenger astronomy which also includes the analysis of electromagnetic radiation and particles physics. 
Neutron Stars (NS) are known to have important magnetic fields \cite{Sinha:2010fm}. Magnetic pressure due to Lorentz force for instance, can distort the star if the magnetic axis is not aligned with the rotation axis. This model is frequently used to explain the pulsar phenomenon \cite{1994ApJ...422..671A}.
Highly magnetized rotating NS are characterized by spin periods typically between $P\sim (10^{-3}-12)s$ and with a change in time of $\frac{dP}{dt}\sim (10^{-16}-10^{-12}) s s^{-1}$ \cite{Vigano:2013uia, Kaspi:2017zrx}. These approximations are consistent with magnetic fields on the surface of pulsars of the order of $B\sim10^{11}-10^{13}$G \cite{Woods:2004kb,Duncan:1992hi}, or as high as $B\sim10^{14}-10^{15}$G in magnetars \cite{Kaspi:2017fwg}, and even up to $10^{16}$G are values likely to be found in new born NS \cite{Ioka:2000yb}. 

Different scenarios have been investigated in the literature where magnetic fields play a major role, for instance, magneto-hydrodynamic simulations of the coalescence of a binary NS performed in \cite{Price:2006fi} 
showed that magnetic fields of NS progenitors of $10^{12}$G are amplified by several orders of magnitude within the first millisecond after the merger via Kelvin-Helmoltz instabilities.

Merging  BNS are also thought to be progenitors of short gamma-ray bursts due to the strong magnetic field of one or both binary members and their high orbital frequencies \cite{Troja:2010zm, Lipunov:1996wf, Hansen:2000am, Medvedev:2012qf, Paschalidis:2013jsa}.
Other studies suggest that the magnetic interaction between two coalescing neutron stars modify considerably the waveform of gravitational radiation \cite{Giacomazzo:2011ilm,Antoniadis:2013pzd,Sarin:2020gxb}.
The fully  general relativistic, magneto-hydrodynamics simulations performed in \cite{Palenzuela:2013hu, Palenzuela:2013kra} have shown that some BNS mergers are ideal candidates for multimessenger astronomy due to their distinctive angular and time dependent pattern in the Poynting flux and their strong emission of GWs.

BNS are characterized by intrinsic parameters such as the mass and spin of the components and extrinsic parameters as the position at the sky, the inclination of the orbital axis and the luminosity distance. However, only intrinsic parameters are directly related to the dynamics of the binary and the shape of the gravitational waveform. 

Despite the fact that current GWs observations cannot be used to describe magnetic field interactions between NS in a binary system, the number of observations of BNS mergers will increase\footnote{For BNS, the median sky localization area of LVC during its third run will be on the order of $10^{5}$ Mpc$^3$, where is expected to detect $1^{+12}_{-1}$ BNS mergers in a year observing run \cite{KAGRA:2013rdx}.}, as well as the sensitivity of the detectors \cite{ligo20}. Thus, the inclusion of magnetic fields in the description of NS  mergers becomes important. In this work, we show that strong magnetic fields change the estimation of the intrinsic parameters of the binary; computing that the change is small but the effect can be inferred from the GW signal.
	
GWs templates have been model in the inspiral phase, where two objects are orbiting and approaching to each other with the orbital frequency increasing. At this stage, the post-Newtonian approach to general relativity has proved to be an adequate model for the evolution of the system and high accurate signals can be computed \cite{Blanchet:1995}. 

In this work, we consider an isolated magnetized two-body system in the linear regime of general relativity; where the Newtonian gravity continues adequately modeling the inspiral phase, and estimate the GW emission through the quadrupolar formalism\footnote{It has been shown that, under certain given circumstances, several effects could modify the waveforms emmited by the binary at 2PN order, 
%in the same amount as the Newtonian quadrupole formalism describes the inspiral stage \cite{Shapiro:1983du,Maggiore:2007ulw}, 
among them are: the post-Newtonian \cite{Blanchet:2006xj} spin-orbit \cite{Gergely:1998vp}, spin-spin \cite{Gergely:1999pd}, self spin, even a quadrupole-monopole coupling \cite{Gergely:2002fd,Vasuth:2003yr}. Furthermore, tidal contributions \cite{Dietrich:2018uni,Bernuzzi:2014owa} have been development and these contributions might affect the dynamics. In the present work we concentrate on the magnetic dipole-dipole effect \cite{Ioka:2000yb} and consider the case where the effect on the dynamics of the rotation of the stars are negligible.}
\cite{Maggiore:2007ulw}.
Newtonian dynamics during the inspiral phase has proved to be a valid approximation as long as the orbital separation of the binary is larger than a minimum radius and the orbital velocity is smaller than the velocity of light.
Following the approach introduced in Ref. \cite{Ioka:2000yb}, we show that using a dipole model for the NS, the dynamics of the binary reduces to an effective one-body problem for the reduced mass, very similar to the non magnetized case. With this model we compute the effect of the magnetic field on the BNS dynamics in a closed form. Furthermore, we compute the GW strains for some values of the magnetic field and use them to estimate the mass of the individual stars.

We apply the model of magnetized stars to two different systems: the first one considers fiducial values for a BNS and explore the effect of the magnetic field in several observables, such as the time to reach a minimal radius, the change in the orbital period, the gravitational luminosity, the logarithmic change in the GW frequency and the strain. We obtain that the deviation in these variables, with respect to the un-magnetized case, can be up to O$(10^{-5})$, when the strenght of the magnetic fields is $B=8\times 10^{16}$ G. In the second approach, using the gravitational luminosity and the logarithmic change in the orbital period as known functions, we determine how the inferred binary masses change as a function of the magnetic field present. We obtain that, with strength of magnetic of the order of $B \sim10^{16}$G, 
that the mass can be sub-estimated or over-estimated with respect to the non magnetized case by a factor up to $4\,\%$. In our analysis, we also consider some astrophysical results of the event GW170817, due its importance as the first evidence of the collision of two NS and its relevance in ongoing astrophysical
developments. Our results are consistent with the magnitude of change in the observable quantities reported in the general relativistic magneto hydro-dynamical simulations, \cite{Anderson:2008zp} providing us with a simple tool to generate accurate templates of waveforms in the last stage of BNS mergers were the magnetic interaction is present.

%\juanc{The following paragraph should be rewritten once the section of results is finished}

%From reported masses of two famous BNS detections, we obtain the deviation caused by an hypothetical magnetic interaction over the orbital frequency, GW frequency, the strain and the gravitational luminosity; all these variables are evaluated at the time to initial separation to the ISCO.  

The paper is organized as follows. In section \ref{sec:setup} we introduce the model for BNS including magnetic fields in the dipole approach and the dynamics of the binary using the Newtonian gravity to describe the orbits. In section \ref{sec:grav_emission} we describe the quadrupolar GW formulation to obtain the GW strain and the mass estimation. In section \ref{sec:results} we present some of the results concerning to the parameter estimation. 
In \ref{sec:res_inspiral} we describe the dynamics and GW emission of two BNS with total mass 2.8 $M_{\odot}$ and fixed magnetic field strengths. %$B_1=B_2=8\times 10^{16}$ G. 
In \ref{sec:res_bvsX} we analyse qualitatively and quantitatively the effect of the presence of magnetic dipole moments with magnetic field strength ranging from $B \sim 10^{12}$G to $B \sim 10^{16}$G on some relevant inspiral variables. This strength lies well into the range of systems described in references
\cite{Kaspi:2017zrx,Kaspi:2017fwg}. In \ref{sub:mass_estimation},
we use the gravitational luminosity and the logarithmic rate of change of orbital period of the event GW170817
to show that
given the uncertainly in the mass determination from observational data, it is possible to get a bound for the maximum value of the magnetic field strength such that the mass
derived from the magnetized case is still consistent with the observations. Finally, in section \ref{sec:conclusions} we give some conclusions.

In this paper we use cgs-gaussian units, where the magnetic fields is measured in Gauss (G = g$^{1/2}$ cm$^{-1/2}$ s$^{-1}$) and the magnetic moments in emu = g$^{1/2}$ cm$^{5/2}$ s$^{-1}$. 

%%%%%%%%%%%%%%%%%%%%%%%%%%%%%%%%%%%%%%%%%%%%%%%%
\section{SET UP: Magnetized binary system}
\label{sec:setup}
%%%%%%%%%%%%%%%%%%%%%%%%%%%%%%%%%%%%%%%%%%%%%%%%

In section \ref{sec:assumptions} we review the conditions under which the gravito-magnetic potential can be analyzed as an one-body problem in a BNS system. Once the magneto-gravitational interaction is described as a central potential, in section \ref{sec:newt} we develop the equations of motion from the Lagrangian of the BNS.  The main purpose of this section is to explore the magnetic effect in the BNS classical description, as a necessary tool to incorporate the effects of the magnetic field in the GW emission in the inspiral analysis,  section \ref{sec:grav_emission}.  

%%%%%%%%%%%%%%%%%%%%%%%%%%%%%%%%%%%%%%%%%%%%%%%%%%%
\subsection{Neutron star model}\label{sec:assumptions}
%%%%%%%%%%%%%%%%%%%%%%%%%%%%%%%%%%%%%%%%%%%%%%%%%%%

Analyzing a magnetized BNS system in the Newtonian regime, we assume that each object is characterized intrinsically by its mass $M_1$ and $M_2$ and its magnetic dipole moment $\mathbf{m}_1$, $\mathbf{m}_2$. 
%$M=M_1+M_2$ is the total mass of the system. 
We isolate the magnetic effect neglecting other NS properties. 
The positions of the NS, $\mathbf{r}_i(t)$, are determined in a reference frame with origin in the center of mass of the system. In this frame the condition $M_1 \mathbf{r}_1 +M_2 \mathbf{r}_2=0$ follows by definition. This reference frame has the advantage that the dynamics reduces to the one-body description with reduced mass $\mu=M_1M_2/M$, and position $\mathbf{r}\equiv \mathbf{r}_1-\mathbf{r}_2$. 
$M=M_1+M_2$ is the total mass of the system.
In this work, we use the magnetostatic approximation for the interior of the stars 
considering that 
the net electric charge is zero \cite{Gruzinov:2007qa,Cerutti:2016ttn}. We also assume that the external magnetic field of each star is a perfect dipole field ${\bf m}_i$
%=R_i^3/2 {\bf B}_i$ 
and consequently, the magnitude of the magnetic moment is only related to the radius of the star as described in \cite{Ioka:2000yb}\footnote{The radius of each star is related to their mass through an equation of state (EoS) \cite{Lattimer:2012nd}. A realistic EoS of NS matter is not known precisely yet, but for masses from 1 to $2M_{\odot}$ \cite{Demorest:2010bx,Antoniadis:2013pzd} different EoS give radii varying from 8 to 16 km \cite{Palenzuela:2013hu,1997ApJ...476L..39C}. 
%For that reason we not consider 
%the radius of each star as intrinsic parameters. 
For simplicity, we use a typical  value for the radius of the stars as $12$ km.}. 

In the magnetostatic regime, the magnetic field ${\bf B}_1$ due to  star 1 at any point 
$\mathbf{x}$, is given by:
\begin{equation}\label{eq:b1}
{\bf B}_1({\bf x})=\frac{3\hat{{\bf n}}_1 \left(\hat{{\bf n}}_1 \cdot {\bf m}_1 \right)-{\bf m}_1}{|{\bf x}-{\bf r}_1|^3} \, ,    
\end{equation}
where $\hat{{\bf n}}_1=({\bf x}-{\bf r}_1)/|{\bf x}-{\bf r}_1|$.

The magnetic potential energy resulting from the interaction of a magnetic dipole ${\bf m}$ with an external magnetic field ${\bf B}$ is given by the dot product $U_m = - {\bf m} \cdot {\bf B} $  \cite{Jackson:1998nia}.

The magnetic potential energy at position $ {\bf r}_2$, is thus
\begin{equation} 
U_m(1 \rightarrow 2)=-{\bf m}_2\cdot{\bf B}_1({\bf r}_2)=
-{\bf m}_2\cdot\frac{3\hat{{\bf r}} \left(\hat{{\bf r}} \cdot {\bf m}_1 \right)-{\bf m}_1}{|{\bf r}|^3}=- \frac{3({\bf m}_2 \cdot \hat{{\bf r}}) \left(\hat{{\bf r}} \cdot {\bf m}_1 \right)-{\bf m}_2 \cdot {\bf m}_1}{|{\bf r}|^3}, \label{U12}
\end{equation}
where we have used Eq. \eqref{eq:b1} and the unit vector $\hat{{\bf r}}=({\bf r}_2-{\bf r}_1)/|{\bf r}_2-{\bf r}_1|$.  

Following \cite{Anderson:2008zp, Liu:2008xy} we assume 
the magnetic moments remain parallel to the total angular momentum
$ \mathbf{L}=\mu\, \left(\mathbf{r} \times \dot{\mathbf{r}}\right)$ during the inspiral, thus 
${\bf m}_1 \cdot \hat{\bf r}={\bf m}_2 \cdot \hat{\bf r}$=0, consequently the magnetic torque between the dipoles $\mathbf{N}=\mathbf{m}_1 \times \mathbf{B}_2$ vanishes  \cite{Jackson:1998nia} and Eq. \eqref{U12} become
\begin{equation}
U_m(1 \rightarrow 2)= \frac{{\bf m}_2 \cdot {\bf m}_1}{|{\bf r}|^3} \label{poten} \ .
\end{equation}
Furthermore we will assume the magnetic moments for each star are of the form  $ m_i =R_i^3\,  B_i/2$, thus we introduce a \textit{magnetic parameter} $b$ as the dot product of the magnetic moments: 
\begin{equation}\label{eq:b}
b \equiv {\bf m}_1 \cdot {\bf m}_2 
= \pm \frac{(R_1 R_2)^3 B_1 B_2}{4}\, ,
\end{equation}
where $+$ or $-$ indicates if the dipoles are aligned or anti-aligned.
Using Eq. \eqref{eq:b} we show that it is possible to encode the magnetic interaction between the dipoles of the binary thorough the magnetic potential energy of the form $U_m =b/r^3$. 

%Also, we assume that the magnetic moments are parallel to the orbital angular momentum $\bf{L}$ given by $\mathbf{L}=\mu\, \left(\mathbf{r} \times \dot{\mathbf{r}}\right)$, i.e., ${\bf L} \times {\bf m}_i =0$. The definition of the orbital angular momentum  and our restriction means that ${\bf m}_i$ are orthogonal to $\bf{r}$ and $\dot{\bf{r}}$ all the time i.e. ${\bf m}_i \cdot {\bf r}=0$. With this, the j-magnetic field in $i$-position is reduced to ${\bf B}_j(r)=-\mathbf{m}_j/r^3$, which means that ${\bf m}_j$ is anti-parallel to ${\bf B}_j$. So, although in general the dipole-dipole interaction generates a magnetic torque on the $i$-object given by $\mathbf{N}_{i}=\mathbf{m}_i \times \mathbf{B}_j$  \cite{Jackson:1998nia}, our restrictions also implies that the torques are zero since ${\bf m}_i$ is parallel to ${\bf B}_j$. 
%
%Besides, the magnetic potential $U_m=- \mathbf{m}_i\cdot\mathbf{B}_j$  reduces to a central potential function of the distance between the NS, $U_m =b/r^3$, where $b$ is given by Eq. (\ref{eq:b}). 
%We see that aligned magnetic moments ($b<0$) implies a repulsive magnetic potential and anti-aligned magnetic moments ($b<0$) an attractive magnetic potential.  

At this point, we can estimate, in orders of magnitude, the fraction of the magnetic potential respect to the gravitational contribution $U_g=-GM\mu/r$, as  $U_m/ U_g=-b/(GM\mu r^2)$ or
\begin{equation}\label{eq:umg} 
\frac{U_m}{U_g}= 1.0\times 10^{-4} \left(\frac{R_1}{12\, \mathrm{km}}\right)^3 
\left(\frac{B_1}{10^{16}\,\mathrm{G}}\right)
\left(\frac{R_2}{12\, \mathrm{km}}\right)^3
\left(\frac{B_2}{10^{16}\,\mathrm{G}}\right) \left(\frac{2.8\,M_{\odot}}{M}\right) \left(\frac{0.7\,M_{\odot}}{\mu}\right) \left(\frac{12\,\mathrm{km}}{r}\right)^2,
\end{equation} 
where we have used the Eq. (\ref{eq:b}) and re-scaled the parameters of BNS in terms of canonical NS with equal masses  $\sim1.4\,M_{\odot}$ and radii $\sim12$ km. For the magnetic field strength we use $\sim 10^{16}$G because it is the maximum estimated for magnetars (but not theorized). 

We can observe from Eq. (\ref{eq:umg}), that the contribution of magnetic over the gravitational potential is ${\cal O}(10^{-4})$. It is in accordance with the post-Newtonian analyzes, which reports that the magnetic dipole-dipole interaction produces a 2PN correction \cite{Ioka:2000yb,Mikoczi:2005dn,Gruzinov:2007qa}. 

%%%%%%%%%%%%%%%%%%%%%%%%%%%%%%%%%%%%%%%%%%%%
\subsection{Newtonian dynamics} \label{sec:newt}   
%%%%%%%%%%%%%%%%%%%%%%%%%%%%%%%%%%%%%%%%%%%%

As described above, the magnetic interaction between the dipoles of the stars in the binary can be described using a central potential. Thus the gravito-magnetic interaction, in the center of mass frame,
is given by the sum of the magnetic $U_m$ and gravitational $U_g$ potential:
\begin{equation}\label{eq:U}
U(r) =  -\frac{G M\mu}{r} +\frac{b}{r^3}=-\frac{GM\mu}{r}\left(1-\frac{b}{GM\mu r^2}\right) \ ,
\end{equation}
where $b$ has been defined in Eq. (\ref{eq:b}). 
With this potential we can use the lagrangian formalism to 
reduce the dynamics of the binary to an equivalent one-body problem for the reduced mass $\mu$ located at ${\bf r}$.
The position of each star is recover using the relations
${\bf r}_1=(M_2/M)\,{\bf r}$ and ${\bf r}_2=-(M_1/M)\,{\bf r}$.
Since the potential $U$ depends only on the position $r$, the total angular momentum is conserved and thus the movement is restricted to a plane. For simplicity, we choose the equatorial plane.
The Lagrangian for the BNS system is thus defined as the difference of the kinetic energy $T=\mu ~ \dot{r}^2/2 +\mu r^2\dot{\varphi}^2/2$, and the potential energy given in Eq. (\ref{eq:U}),
%Therefore the Lagrangian of the BNS is given by
\begin{equation}\label{eq:lag}
\mathcal{L}(r,\varphi)=\frac{\mu}{2}\dot{r}^2 + \frac{\mu}{2} r^2\dot{\varphi}^2 +\frac{G M\mu}{r}-\frac{b}{ r^3} \,.
\end{equation}
Since the Lagrangian in Eq. \eqref{eq:lag} is independent of $\varphi$, one obtains the conservation of the angular momentum directly from the Euler-Lagrange equations 
\begin{equation}\label{eq:eul_phi}
\frac{d}{dt} (\mu r^{2} \dot{\varphi})=0 \hspace{.3cm}\Rightarrow\hspace{.3cm}l=\mu r^2\dot{\varphi}={\rm const} \ .
\end{equation}
The total energy of the system $E=T+U$ can be written as 
\begin{equation}
\label{eq:E_Veff}
E=\frac{1}{2}\mu \dot{r}^2 + V_{\rm eff}(r) \ ,
\end{equation}
where $V_{\rm eff}(r)\equiv \frac{l^2}{2\mu r^2} - \frac{GM\mu}{r} + \frac{b}{r^3}$ is an effective potential and we have used the relation between $l$ and $\dot \varphi$ given in Eq. (\ref{eq:eul_phi}).
On the other hand, the Euler-Lagrange equation for the coordinate $r$ is
\begin{equation}\label{eq:eul_r}
\ddot{r}+\frac{GM}{r^2}-\frac{l^2}{\mu^2 r^3}-\frac{3b}{\mu r^4}=0 \ .
\end{equation}
Setting a variable change $u=1/r$ we have
\begin{equation}\label{eq:2ordu}
\dot r = -\frac{1}{u^2}\dot u= -\frac{l}{\mu} \frac{\dot u}{\dot \varphi}  = -\frac{l}{\mu} \frac{du}{d\varphi} \qquad {\rm and } \qquad  \ddot r  =
-\frac{l}{\mu}\dot \varphi\frac{d^2u}{d\varphi^2}
=
-\frac{l^2}{\mu^2} u^2\frac{d^2u}{d\varphi^2} \ ,
\end{equation}
so that, after some simplifications, Eq. \eqref{eq:eul_r} becomes
\begin{equation}\label{eq:orb}
\frac{d^2 u}{d\varphi^2} + u - \frac{1}{R}= \delta_b u^2\ ,
\end{equation}
where 
\begin{equation} 
R \equiv \frac{l^2}{GM\mu^2} \qquad {\rm and} \qquad  \delta_b = - \frac{3\mu b}{l^2}    \ . \label{eq:R}
\end{equation}
The right hand side of Eq. \eqref{eq:orb} is a nonlinear term induced by the dipole interaction. 
The solutions with no magnetic field $b=0=\delta_b$, are the conic sections $u(\varphi)=\frac{1}{R} \left( 1+\epsilon \cos\varphi \right) $,
with 
eccentricity $\epsilon^2\equiv 1+2El^2/(G^2M^2 \mu^3)$. %
Notice that the nonlinear term in Eq. \eqref{eq:orb} has the same form as the relativistic correction to the Newtonian potential given by the Schwarzschild spacetime in a relativistic treatment. In the following section we show that circular orbits are allowed for a range of values of the magnetic strength.

%%%%%%%%%%%%%%%%%%%%%%%%%%%%%%
\subsection{Circular motion}
%%%%%%%%%%%%%%%%%%%%%%%%%%%%%%

It is know that the emission of gravitational radiation tends to circularize elliptical orbits to the degree that before  merger, the orbits have been circularized \cite{Lincoln:1990ji,Blanchet:1996pi,Blanchet:2013haa}. Since the circular motion dominates the BNS dynamics during the inspiral phase we focus our analysis for this type of orbits.

\begin{figure}[!ht]
	\begin{centering}
		\includegraphics[scale=.5]{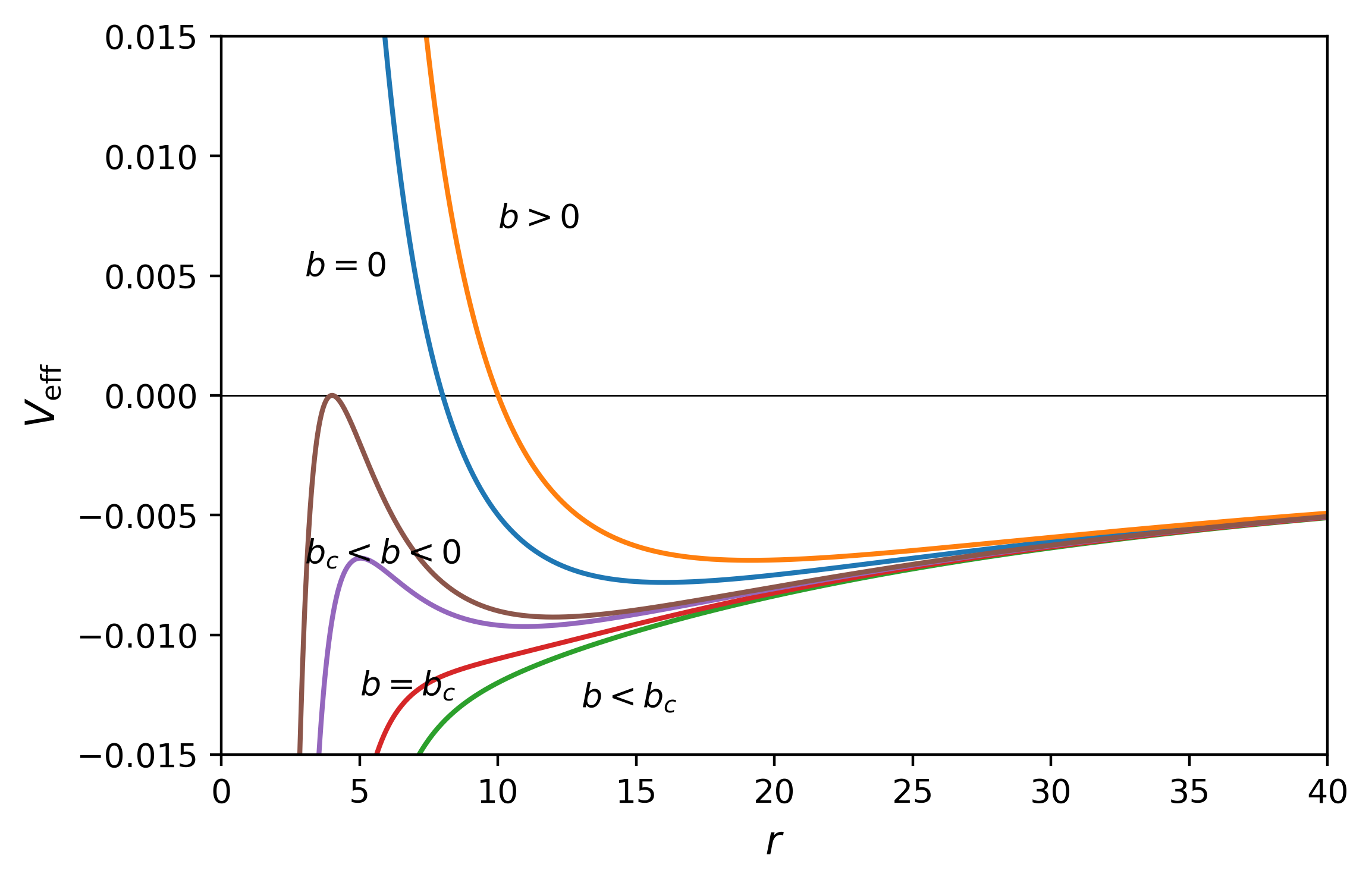}
		\par
	\end{centering}
	\caption{Effective potential $V_{\rm eff}$. For no-negative values of $b$ there is always an stable circular orbit. For $b_c<b<0$ an unstable circular orbit appears. The critical value $b_c=-\frac{l^4}{12GM\mu^3}$ 
	represents the last circular orbit. For more negative values of $b$ there are no circular orbits. }
	\label{fig:Veff}
\end{figure}

Circular orbits ($r=$ cte.) may be possible if the condition $\ddot r =0$ in Eq. \eqref{eq:eul_r} is satisfied. In the scenario described in this work, it happens to be the case for a combination of the magnetic field strength and the angular momentum.
By setting $\ddot r=0$ in Eq. \eqref{eq:eul_r} 
and solving for $r$ we obtain
\begin{equation}\label{eq:rcir}
    r_{\rm c} = \frac{R}{2}\left(1 + \sqrt{1+\frac{12\,b}{GM\mu\,R^2}}\right) \ ,
\end{equation}
where we have used $R$ defined in Eq. \eqref{eq:R}. For $b>0$ there is always a circular orbit; for $b<0$ there is a  critical value $b_{c}=-\frac{l^4}{12GM\mu^3}$ below of which circular orbits ceases to exit. For more negative values of $b$ the 
effective potential has not extreme points.
Fig. \ref{fig:Veff} displays $V_{\rm eff}$ for some representative values of $b$.
The angular momentum 
$l$ for circular orbits is expressed by:  
\begin{equation}\label{eq:lcr}
l_b = \mu \sqrt{G\,M\,r_c \left(1-\frac{3\,b}{G\,M\, \mu\,r_c^{2}}\right)} \ ,
\end{equation}
where subscript $b$ denotes a dependence of the variable to the magnetic parameter $b$.  Moreover, from Eq. (\ref{eq:eul_phi}), $\dot\varphi=l/\mu r^2$, the orbital frequency in circular orbits is given by
\begin{equation}\label{eq:dphi}
\dot\varphi_b = \sqrt{\frac{G\,M}{r_c^3} \left(1-\frac{3\,b}{G\,M\, \mu\,r_c^{2}}\right)}\ ,
\end{equation}
this expression is the analogous to the third Kepler law for circular orbits and will be use in the next section. 

Finally, the total energy of circular orbits is 
\begin{equation}\label{eq:Ec}
E_b = - \frac{G\,M\,\mu}{2r_c}\,\left(1 + \frac{b}{G\,M\,\mu\,r_c^2}\right) \ , 
\end{equation}
which is equal to the minimum of the effective potential $V_{\rm eff}$ in Eq. (\ref{eq:E_Veff}).

%%%%%%%%%%%%%%%%%%%%%%%%%%%%%%%%%%%%%%%%%%%%%%%%%%%%%
\subsection{Orbit precession due to nonzero magnetic field}
%%%%%%%%%%%%%%%%%%%%%%%%%%%%%%%%%%%%%%%%%%%%%%%%%%%%

A slightly non-circular orbit will oscillate in and out about a central radius. In the absence of magnetic field the allowed bounded orbits are ellipses. 
The presence of the magnetic term modifies this behaviour;
the orbit looks like an ellipse which slowly rotates about the center, this phenomenon is known as the precession in the orbit.

Eq. \eqref{eq:2ordu} can be solved numerically quite straightforwardly, however, some useful information can be inferred from the solution in the limit of small magnetic interaction. Assuming $\delta_b<<R$, we seek approximate solutions of Eq.  \eqref{eq:2ordu} of the form
\begin{equation} \label{eq:usum}
u= u_0 + \beta u_1 \ ,   \qquad {\rm with} \qquad \beta = \frac{\delta_b}{R} \ ,
\end{equation}
were we have neglected higher powers of $\delta_b/R$. Substituting Eq. \eqref{eq:usum} into Eq. \eqref{eq:orb} and collecting terms of the same order, one gets 
\begin{eqnarray}
\frac{d^2 (R\, u _0)}{d\varphi^2} + R\,u_0  &=& 1, \\
\frac{d^2 (R\, u _1)}{d\varphi^2} + R\,u_1 &=&  ( R\,u_0)^2 \ \label{second} .  
\end{eqnarray}
The solution of the first equation, as mentioned above, is the conic section
\begin{equation} \label{eq:sol_p}
R\,u_0 = 1+ \epsilon \cos \varphi \ ,    
\end{equation}
where $\epsilon^2 \equiv 1+\frac{2El^2}{G^2 M^2\mu^3}$. On the other hand
a particular solution of Eq. \eqref{second} is 
\begin{equation}
    R\,u_1 =  \left( 1 + \frac{\epsilon^2}{2} + \epsilon \varphi \sin\varphi - \frac{\epsilon^2}{6} \cos(2\varphi)\right) \ .
\end{equation}
Notice the third term increases with each orbit and becomes the most relevant. Neglecting the other corrections we can thus write Eq. \eqref{eq:usum}, in the limit $\beta<<1$, as
\begin{eqnarray}
Ru &\approx&  1 +  \epsilon \cos \varphi + \beta\epsilon \varphi \sin\varphi   \nonumber\\
&\approx& 1 + \epsilon \cos \left( \varphi - \beta \varphi \right) \ . 
\end{eqnarray}
Thus, the period of the orbit is no longer $2\pi$ but rather
\begin{equation}
    \frac{2\pi}{1-\beta} \approx 2\pi (1+\beta) = 2\pi \left( 1+ \frac{\delta_b}{R}\right) \ .
\end{equation}
The precession, in radians per orbit, is therefore given by
\begin{equation}
    \Delta \varphi = 2\pi - 2\pi \left( 1+ \frac{\delta_b}{R}\right) =
    2\pi\frac{\delta_b}{R} \ .
\end{equation}
Substituting the expressions of $R$ and $\delta_b$ given by Eq.~\eqref{eq:R} and the expression of $b$ in terms of the strength of the magnetic field, we obtain the precession of the orbit $\Delta\,\varphi = \frac{2\pi\delta_b}{R}=\mp 6 \pi G M \mu^3 R_1^3 R_2^3 B_1 B_2/(4 l^4)$.  
For typical values one gets 
\begin{equation} \label{eq:triangle}
\begin{aligned}
\Delta\,\varphi  = \mp 2.83\times 10^{-7} \left(\frac{R_1}{12\,\mathrm{km}}\right)^3 
\left(\frac{B_1}{10^{16}\,\mathrm{G}}\right)
\left(\frac{R_2}{12\,\mathrm{km}}\right)^3 
\left(\frac{B_2}{10^{16}\,\mathrm{G}}\right) \left(\frac{M}{2.8\,M_{\odot}}\right) \left(\frac{\mu}{0.7\,M_{\odot}}\right)^{3} \left(\frac{l_0}{l}\right)^4. 
\end{aligned}
\end{equation}
where the angular momentum $l$ is obtained from the Keplerian expression $l^2=GM\mu^2a\left(1-\epsilon^2\right)$, for a binary  with 
$a=10^{3}\,{\rm km}$,  
$l_0=2.69\times10^{50}\mathrm{g\, cm}^{2}{\rm s}^{-1}=1.35\times10^7\,M_{\odot}\,{\rm km}^2\,{\rm s}^{-1}$.

%%%%%%%%%%%%%%%%%%%%%%%%%%%%%%%%%%%%%%%%%%%%
In the following section we shall consider the lost of energy of the binary due to GW emission using the quadrupolar formalism. We focus on circular orbits since as mentioned above, GW emission tend to circularize the orbits during the inspiral.

%%%%%%%%%%%%%%%%%%%%%%%%%%%%%%%%%%%%%
\section{GRAVITATIONAL EMISSION}
\label{sec:grav_emission}
%%%%%%%%%%%%%%%%%%%%%%%%%%%%%%%%%%%%%

The linearized theory of gravitation is based on the assumption that the gravitational field is weak and the metric deviates only slightly from a Minkowski spacetime. In many cases GWs may be assumed to represent weak perturbation of the spacetime geometry which permits the metric to be expressed in the form
\begin{eqnarray}
g_{\mu\nu} = \eta_{\mu\nu} + h_{\mu\nu} \ ,
\end{eqnarray}
where $\eta_{\mu\nu}$ is the metric of the flat Minkowski spacetime in cartesian coordinates and $|h_{\mu\nu}| \ll 1$.
In the Lorentz gauge, the linearized Einstein's equations to be
\begin{equation}\label{eq:we}
\square \bar{h}_{\mu\nu}= - \frac{16\pi G}{c^4} T_{\mu\nu} \ , 
\end{equation}
where the trace-reversed amplitude $\bar{h}_{\mu\nu}$ is defined by  $\bar{h}_{\mu\nu}\equiv h_{\mu\nu}-\eta_{\mu\nu}h/2$ and $h= h^{\mu}{}_{\mu}$. 

In the long-wavelength approximation (wavelength of GWs are much larger than the characteristic source size), for far non-relativistic sources, and for a GW propagating in the $z$ direction, the solution of Eq. (\ref{eq:we}) is given in terms of the two polarization amplitudes, $h_+$ and $h_\times$ as in reference \cite{Maggiore:2007ulw},
\begin{equation}\label{eq:h}
h_+(t)=\frac{1}{d}\frac{G}{c^4} \left(\ddot{M}_{xx}(t_r)-\ddot{M}_{yy}(t_r)\right),\hspace{1cm} h_{\times}(t)  = \frac{2}{d} \frac{G}{c^4}\ddot{M}_{xy}(t_r) \ ,
\end{equation}
where $d$ is the distance from the source to the Earth detectors and  $M^{ij}\equiv \int d^3x T^{00}(t,\mathbf{x})x^ix^j$ is the second mass moment (or quadrupole symmetric tensor).
The dots represents time derivatives with respect to $t_r$ and $t_r \equiv t-d/c$ is the retarded time.

The change of gravitational luminosity per unit solid angle is related to the polarization amplitudes as 
\begin{equation}
\frac{dL}{d\Omega}=\frac{r^2 c^3}{16 \pi G} \langle \dot{h}^2_+ + \dot{h}^2_{\times} \rangle, \label{luminosity}   
\end{equation} 
where $L$ is named the gravitational luminosity. Inserting Eq. \eqref{eq:h} in Eq. \eqref{luminosity}, we find the total radiated power
\begin{equation}\label{eq:L}
L = \frac{G}{5c^5} \left \langle \dddot{M}_{ij}\dddot{M}_{ij} - \frac{1}{3} (\dddot{M}_{kk})^2  \right \rangle,
\end{equation}
where the average is a temporal average over several characteristic periods of the GWs.
Considering the binary system (including magnetic field) described in section \ref{sec:setup},
where $z$ axis is perpendicular to the plane of motion and $\varphi_b$ is the angle from the $x$ axis to the line joining the masses, the second mass moment of the system is
\begin{equation}\label{M}
M^{ij}_b(t)=\frac{1}{2}\mu r^2\left(\begin{array}{ccc}
1+\cos(2\varphi_b(t)) & \sin (2\varphi_b(t)) & 0\\
\sin (2\varphi_b(t)) & 1-\cos (2\varphi_b(t)) & 0\\
0 & 0 & 0
\end{array}\right) \ . 
\end{equation} 
For circular orbits including magnetic field (with $\dot{r}=0$ and $\ddot{r}=0$), the gravitational luminosity is 
\begin{equation}\label{lgm}
\begin{aligned}
L_{b} & =  \frac{32}{5}\frac{G^4M^3\mu^2}{c^5 r^5} \left(1-\frac{3b}{GM\mu r^2}\right)^3 \ ,\\ 
\end{aligned}
\end{equation}
where we used the time dependence of $\varphi_b$ given by Eq. \eqref{eq:dphi}.
The loss of energy through GW emission via the relation $L_b + \dot{E}=0$ leads to a decrease in the separation $r$ as
\begin{equation}\label{eq:dr}
\left(\frac{dr}{dt}\right)_b= - \frac{64 G^3M^2\mu}{5 c^{5}r^{3}} 
\left(1 - \frac{3b}{G M \mu r^{2}}\right)^{3} 
\left(1 + \frac{3b}{G M \mu r^{2}}\right)^{-1},
\end{equation} 
where we have used Eq. \eqref{lgm} and the fact that
$\dot{E}=\frac{1}{2}\left(\frac{GM\mu}{r^2}+\frac{3b}{r^4}\right) \dot{r}$.  
From the previous analysis, it can be seen that when the magnitude of the magnetic parameter $b$ matches the critical value $b_c$ of the Newtonian treatment, there are not circular orbits and the system simply collapses.
Furthermore, as a consequence of the decreasing in $r$, the orbital period $P_b = 2\pi/\dot{\varphi_b}$ also decreases; and the logarithmic rate of change of $P_b$ is
\begin{equation}\label{eq:ppgm} 
\begin{aligned}
\frac{1}{P_b}\frac{dP_b}{dt}  = &  - \frac{96G^3 M^2 \mu}{5 c^5 r^4}\left(1-\frac{3b}{GM\mu r^2}\right)^2 \left(1-\frac{5b}{GM\mu r^2}\right)\left(1+\frac{3b}{GM\mu r^2}\right)^{-1} . \\
\end{aligned}
\end{equation}
Notice that setting $b=0$ in Eqs. (\ref{eq:dr}, \ref{eq:ppgm}) they reduce to the well known expressions given for instance in reference \cite{Shapiro:1983du}. 

%%%%%%%%%%%%%%%%%%%%%%%%%%%%%%%%%%
\subsection{GW strain estimates}
%%%%%%%%%%%%%%%%%%%%%%%%%%%%%%%%%%

The decrease in the separation of the neutron stars, going through a succession of quasi circular orbits, is driven by the emission of GW until the stars merge. However, when the stars are very close to each other, the dynamics is dominated by strong field effects and any approximation of GR used to describe the dynamics ceases to be valid. 

As the two stars rotate around each other their orbital distances decreases, this causes the frequency of the GWs to increase.  However,
the frequency is only valid up to a maximum frequency beyond which the inspiral phase ends and $r_{\rm min}$ is reached.

The time in which the binary system reaches a minimal radius $r_{\rm min}$ can be computed as 
 
\begin{equation}\label{eq:tau}
    \tau_b=\int_{r_0}^{r_{\rm min}} (dr/dt)_b^{-1}\,dr\ ,
\end{equation}
where $r_0$ is the separation between the stars at time $t=0$ and the change of $r$ with time $(dr/dt)_b$ is given by Eq. (\ref{eq:dr}).

Departing from circular orbits and the second mass moment given in Eq. \eqref{M} and using Eq. \eqref{eq:h}, the polarization amplitudes are
\begin{equation} \label{eq:hb}
\begin{aligned}
h_{+_b}(t) & = -\frac{4G^2M\mu}{c^4dr_b(t)} \left(1-\frac{3b}{GM\mu r_b(t)^2}\right) \cos(2\varphi_b(t))\ , \\
h_{{\times}_b}(t) & =-\frac{4G^2M\mu}{c^4dr_b(t)} \left(1-\frac{3b}{GM\mu r_b(t)^2}\right) \sin(2\varphi_b(t)) \ ,\\
\end{aligned}
\end{equation}
where $r_b(t)$ is given by the integration of Eq. \eqref{eq:dr}.
The strain of the GW is defined as $h_b=\sqrt{h_{+_b}^2+h_{{\times}_b}^2}$ and using 
Eq. (\ref{eq:hb}),
%at ISCO 
the strain is given by 
\begin{equation}\label{eq:hbisco}
h_b(t)%^{(\rm ISCO)}
=\frac{4G^2M\mu}{c^4\,d r_b(t)} \left(1-\frac{3b}{GM\mu r_b(t)^2}\right).
\end{equation}

Notice that setting $b=0$ in Eqs. (\ref{eq:dr}, \ref{eq:ppgm}) they reduce to the well known expressions given for instance in reference \cite{Shapiro:1983du}. 

For GWs, the term $2\varphi_b$ in Eq. \eqref{eq:hb} can be approximated as $2\varphi_b\simeq 2\dot{\varphi}_b t$ where the GW frequency,
$\omega_{\rm GW}  = 2\dot{\varphi}_b=2\omega_b$ can be obtained. The frequency measured in Hertz is simply $\nu_b=\frac{\omega_{GW}}{2 \pi}$.
number of cycles during the inspiral phase can thus be computed as 
\begin{equation}\label{eq:Ncyc}
\mathcal{N}_{b}=\int_0^{\tau_b} \,\nu_b (t)\,dt \ .
\end{equation}
The derivative of the orbital frequency, $\dot \omega_b$, can be constructed by the chain's rule: $\dot \omega_b = \left(\frac{d\omega_b}{dr}\right)\left( \frac{dr}{dt} \right)$, where $\left(\frac{d\omega_b}{dr}\right)$ is directly calculated from the Eq. (\ref{eq:dphi}), and $\left( \frac{dr}{dt} \right)$ is substituted from Eq. (\ref{eq:dr}). By this way, we obtain
\begin{equation}\label{eq:omega} 
\begin{aligned}
\dot \omega_b  =   \frac{96G^3 M^2 \mu}{5 c^5 r^4}\sqrt{\frac{G\,M}{r^3} \left(1-\frac{3\,b}{G\,M\, \mu\,r^{2}}\right)} \left(1-\frac{3b}{GM\mu r^2}\right)^2 \left(1-\frac{5b}{GM\mu r^2}\right)\left(1+\frac{3b}{GM\mu r^2}\right)^{-1}.
\end{aligned}
\end{equation}
Finally, considering the circular radius of the non-magnetized BNS in Eq.~(\ref{eq:dphi}), we write the radius as $r=(GM/(\omega_0^2))^{1/3}$, with $\dot{\varphi}_0=\omega_0$, where we are denoting $\omega_0$ as the angular frequency which corresponds to this circular non magnetic case. After some algebraic manipulation one gets 
\begin{equation}
\dot{\omega_b} =  \frac{96\,G^{5/3} M^{2/3}\mu {\omega}_0^{11/3}}{5 c^5} \sqrt{1-3k\,{\omega}_0^{4/3}} \left(1-5k\,{\omega}_0^{4/3}\right) \left(1-3k\,{\omega}_0^{4/3}\right)^2 \left(1+3k\,{\omega}_0^{4/3}\right)^{-1},
\end{equation}
where $k\equiv b/(GM)^{5/3}\mu$.  

%%%%%%%%%%%%%%%%%%%%%%%%%%%%%%
\subsection{Mass estimation}\label{sec:mass_estimation}
%%%%%%%%%%%%%%%%%%%%%%%%%%%%%%%

In this section, we show how the expressions for
the gravitational luminosity Eq. \eqref{lgm}
and  the change in the orbital period Eq. \eqref{eq:ppgm} can be used to determine the total mass of the binary. Furthermore, from this total mass, we obtain a measure of the individual masses.
Let us assume an scenario in which we define the ratio $Q=\frac{\dot P}{P}$, and the gravitational luminosity $L$ are known during the inspiral phase. 

From Eqs.~(\ref{lgm}, \ref{eq:ppgm}) and after some algebra we get the next expression for the total mass 
\begin{equation}\label{eq:Mwb}
M =	 \frac{5Lr^9Q^2c^5}{96G^2} \frac{\sqrt{3}A(Lr^3+Qb)+f_1}{f_2+\sqrt{3} A f_3} \ ,
\end{equation}
and the reduced mass
\begin{equation} \label{eq:muwb}
\mu =	-\frac{48G}{Lr^{11}Q^3c^5}\left(\sqrt{3}A f_4 + f_5\right) \ , 
\end{equation}
where 
\begin{eqnarray}
A&=&\sqrt{3L^2r^6+20QbLr^3+12Q^2b^2}, \nonumber \\ f_1&=&3L^2r^6+13QbLr^3+6Q^2b^2, \nonumber \\ f_2&=&9L^4r^{12}+96L^3r^9Qb+320L^2r^6Q^2b^2+389Lr^3Q^3b^3+144Q^4b^4, \nonumber \\ f_3&=&3L^3r^9+22QbL^2r^6+44Q^2b^2Lr^3+Q^3b^3, \nonumber \\ f_4&=&3L^2r^6+14QbLr^3 + 12Q^2b^2, \nonumber \\ f_5&=&9L^3r^9+72QbL^2r^6+144Q^2b^2Lr^3+72Q^3b^3. 
\end{eqnarray}
For a vanishing magnetic field
$b=0$, we recover the known expressions for the total and reduced mass \cite{Maggiore:2007ulw},
\begin{equation}
\label{eq:M0}
M_0 = \frac{5 Q^{2} c^{5} r^{3}}{288 G^{2} L} \qquad {\rm and} \qquad \mu_0 = - \frac{864 G L^{2}}{5 Q^{3} c^{5} r^{2}} \ .
\end{equation}
In order to better understand, the impact of the magnetic field in the determination of $\mu$ and $M$ we define the variable 
\begin{equation}\label{eq:x}
x\equiv \frac{Qb}{L\,r^3} \ .   
\end{equation}
By writing Eq. (\ref{eq:Mwb}) in terms of $x$, we get $M(x)= M_0 f_M(x)$ where
\begin{equation}\label{eq:cM}
f_M(x) \equiv  \frac{3\left(3+13x+6x^2+\sqrt{3}A_x(1+x)\right)}{\sqrt{3}A_x f_6 + f_7}, 
\end{equation}
with the definitions
\begin{eqnarray}
A_x&=&\sqrt{12x^2+20x+3}, \nonumber \\
f_6&=&3+22x+44x^2+24x^3, \nonumber \\
f_7&=&9+96x+320x^2+384x^3+144x^4.
\end{eqnarray}
The reduced mass in terms of the parameter $x$ is obtained from Eq. \eqref{eq:muwb} as
$\mu(x)=\mu_0 f_{\mu}(x)$, where
\begin{equation}\label{eq:cmu}
f_{\mu}(x)\equiv \frac{1}{2} + \sqrt{3}A_x \left(\frac{1}{6}+\frac{7x}{9} + \frac{2x^2}{3}\right) + 4x + 8x^2 +4x^3. 
\end{equation}
According to Eqs. (\ref{eq:Mwb}, \ref{eq:muwb}) for given values of the gravitational luminosity $L$, and the logarithmic change of the GW frequency $Q$, one may deduce a value of the mass $M$ and the reduced mass $\mu$. 
Any deviation from the values in Eqs.~(\ref{eq:M0}) (equivalently if $f_M \neq 1$ and 
$f_{\mu} \neq 1$) 
can thus be associated with a presence of magnetic field. 

\begin{figure}[!ht]
	\begin{centering}
		\includegraphics[scale=.7]{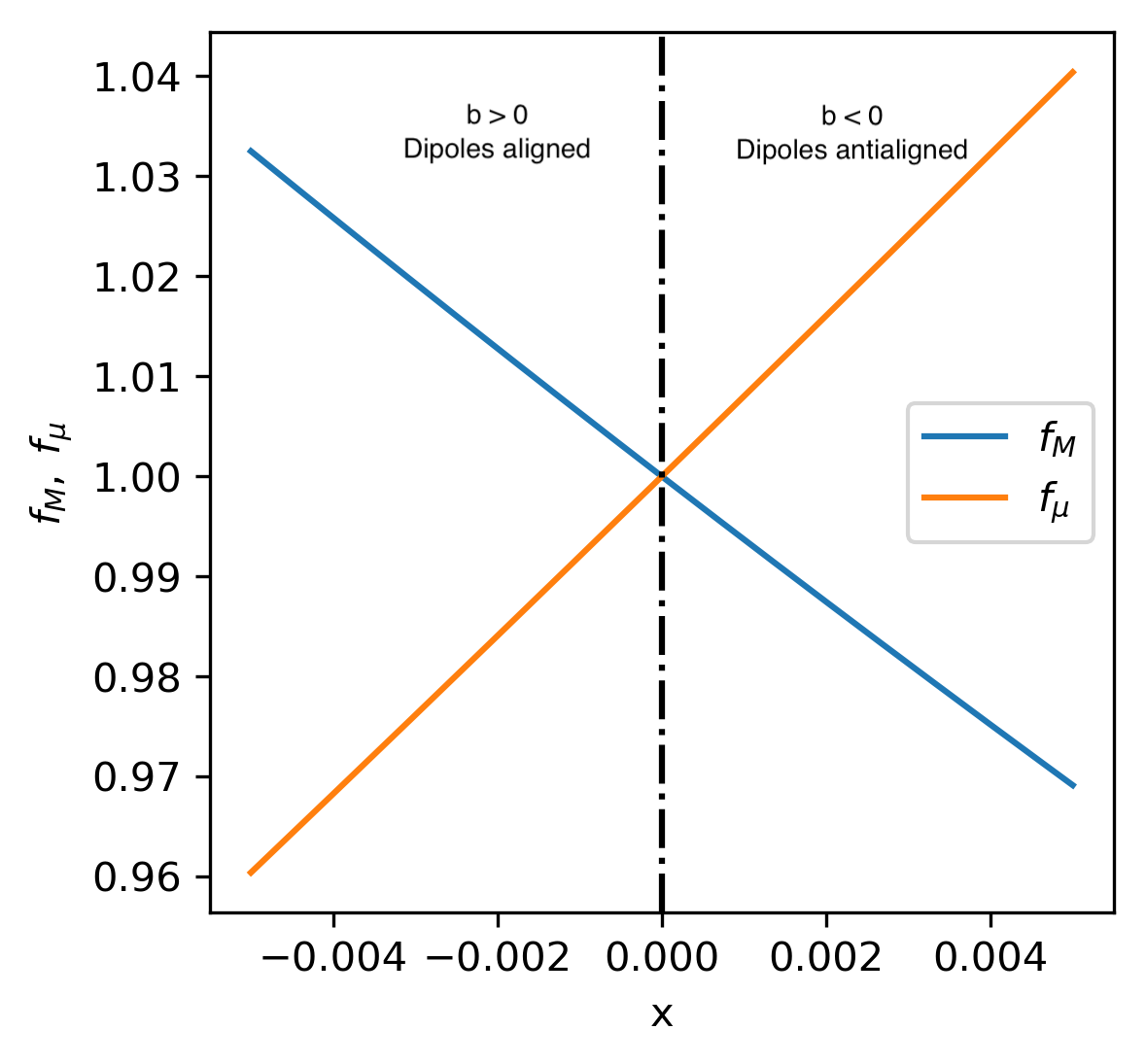}
		\par
	\end{centering}
	\caption{Functions $f_M$ and $f_\mu$ defined in Eqs. (\ref{eq:cM},\ref{eq:cmu}). The intersection $f_M=1=f_\mu$ at the origin, corresponds to the case with no magnetic field.}
	\label{fig:f}
\end{figure}
The determination of the individual masses of the binary can be obtained from the definition of total mass and reduced mass as
\begin{eqnarray}
M_1(x) &=& \frac{1}{2}  M_0f_M(x)  \left(1- \sqrt{  1- \frac{1}{4}\frac  {\mu_0}{M_0} \frac{f_{\mu}(x)}{f_M(x)}   }       \right) \ , \nonumber \\
M_2(x) &=& \frac{1}{2}  M_0f_M(x)  \left( 1+ \sqrt{  1- \frac{1}{4}\frac  {\mu_0}{M_0} \frac{f_{\mu}(x)}{f_M(x)}   }       \right)  \ .
\end{eqnarray}
Notice however,
the dependence of the individual mass depends on  $M_0$ and $\mu_0$ and not on the individual mass of the binary with zero magnetic field.

As we describe in the following section 
a nonzero magnetic field is consistent with
the current uncertainly in the estimation of the mass for systems with equal and non-equal masses. 
Furthermore the maximum and minimum mass estimation for these systems can be used to determine 
a bound for the maximum and minimum value of the allowed magnetic field.
Fig. \ref{fig:f} shows the plot of $f_{M}$ and $f_{\mu}$ in terms of $x$. We focus on a region close to $x=0$ (no magnetic field) since we are interested in the small deviations produced by the magnetic field. Close the origin
the slope of $f_M$ is negative whereas the slope of $f_{\mu}$ is positive and as we show below, this behavior will cause an over or under estimation in the mass of each component in the presence of magnetic field.
Values of $x>0$ correspond to a binary with anti-aligned dipoles and $x<0$ correspond to aligned dipoles. This is consistent with the fact that the anti-aligned configuration storage more potential energy.

In the following section we show the effect of the magnetic field in the gravitational strain and mass estimation using some examples of astrophysical relevance.  

%%%%%%%%%%%%%%%%%%%%%%%%%%%%%%%%%%
\section{Results and Discussion}
\label{sec:results}
%%%%%%%%%%%%%%%%%%%%%%%%%%%%%%%%%%
 
In this section we consider three different approaches in which the formalism developed above can be applied to determine qualitatively and quantitatively the role of the magnetic field in some scenarios of astrophysical interest. 

\iffalse
\mariana{In \ref{sec:res_inspiral} we describe the dynamics and GW emission of two BNS with total mass 2.8 $M_{\odot}$ and fixed magnetic field strengths $B_1=B_2=8\times 10^{16}$ G. 
In \ref{sec:res_bvsX} we analyse qualitatively and quantitatively the effect of the presence of magnetic dipole moments with magnetic field strength ranging from $\sim 10^{12}$G to $\sim 10^{16}$G on some relevant inspiral variables. This strength lies well into the range of systems described in references
\cite{Kaspi:2017zrx,Kaspi:2017fwg}. }
Finally,  in \ref{sub:mass_estimation},
we use the gravitational luminosity and the logarithmic rate of change of orbital period of the event GW170817
to show that
given the uncertainly in the mass determination from observational data, it is possible to get a bound for the maximum value of the magnetic field strength such that the mass
derived from the magnetized case is still consistent with the observations.

\fi

%%%%%%%%%%%%%%%%%%%%%%%%%%%%%%%%%%%%%%%%%%%%%%%%%%%%%%%%%
\subsection{Dipole alignment effect on the BNS dynamics and on the waveform} 
\label{sec:res_inspiral}
\begin{figure}[!ht]
	\begin{centering}
		\includegraphics[scale=.65]{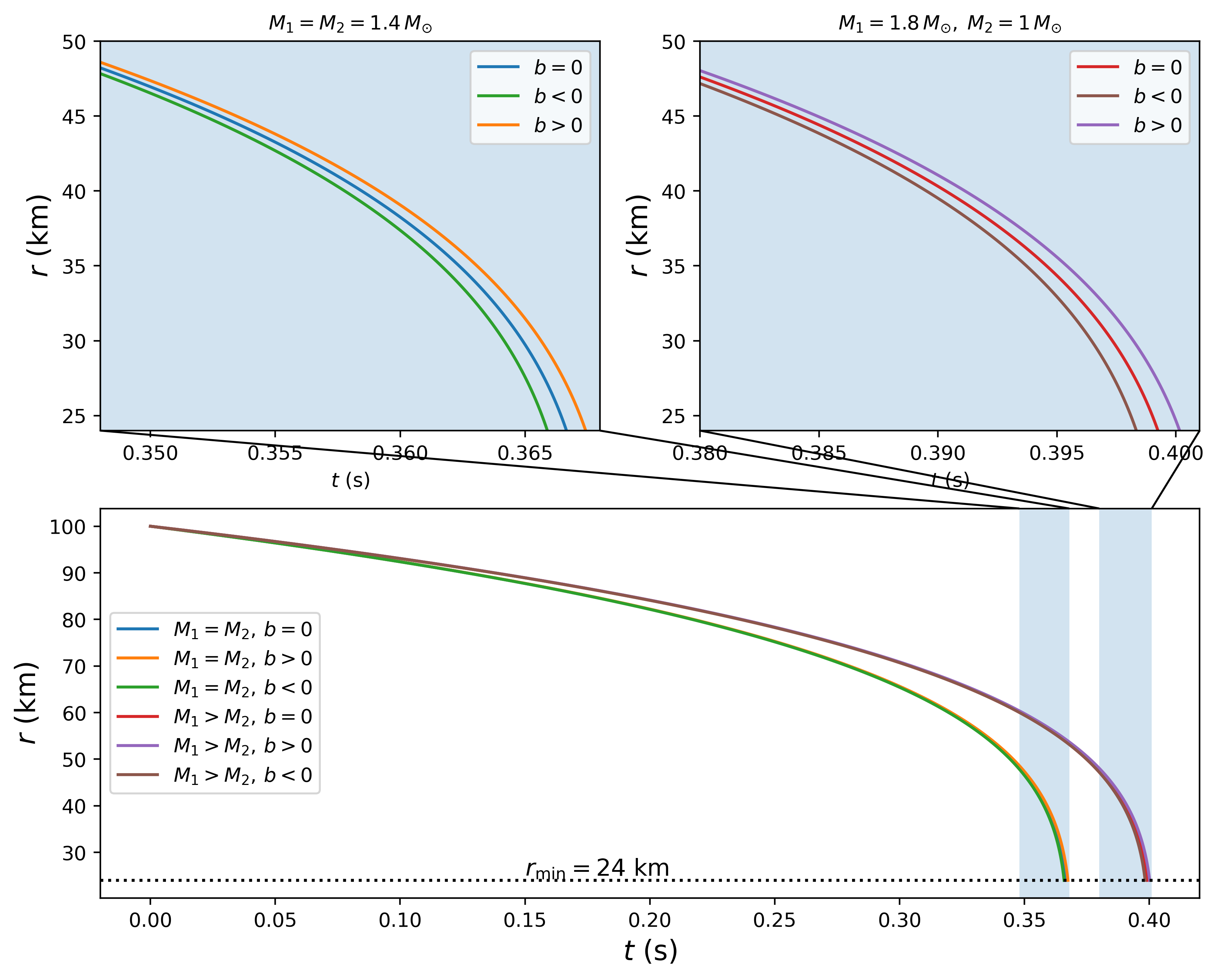} 
		\par
	\end{centering}
	\caption{
	It is presented the evolution of the neutron stars separation for the cases with $M_1=M_2=1.4\,M_\odot$, and $M_1=1.8\,M_\odot, M_2=1\,M_\odot$. The magnetic field of magnitude used to construct these plots is $B=8\times 10^{16}$ G. The top panels represent a 
	closer look  where the magnetic effects are more noticeable near the merger. 
	}
	\label{fig:insp1}
\end{figure}

\begin{figure}[!ht]
	\begin{centering}
		\includegraphics[scale=.65]{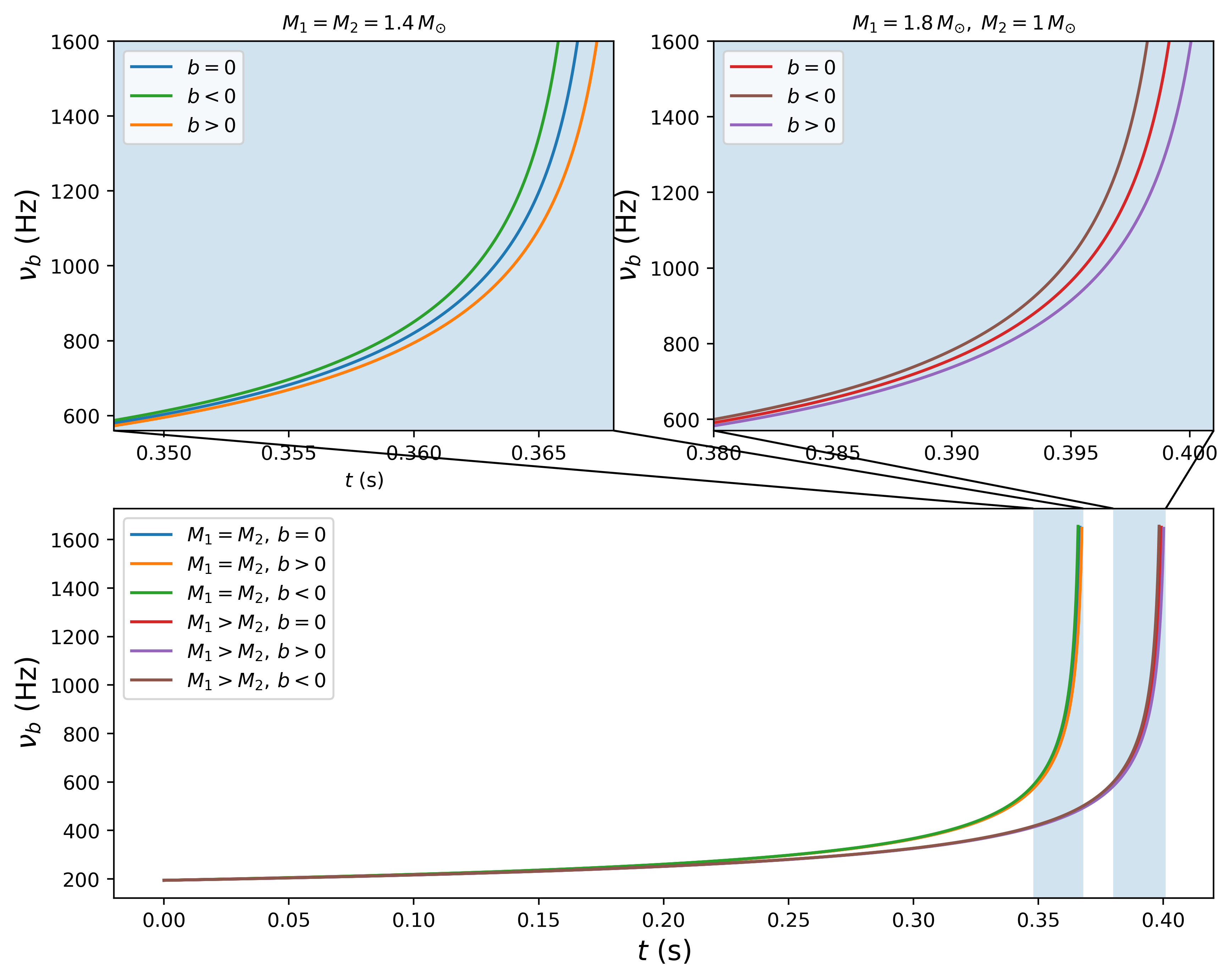} 
		\par
	\end{centering}
	\caption{Evolution of the GW frequency during the last phase of the inspiral for BNS system with equal (top left) and non equal (top right) masses. The time intervals and conditions are same as in Fig. \ref{fig:insp1}. The figure in the bottom shows the six cases during the entire inspiral.
	}
	\label{fig:nub}
\end{figure}

\begin{figure}[!ht]
	\begin{centering}
		\includegraphics[scale=.65]{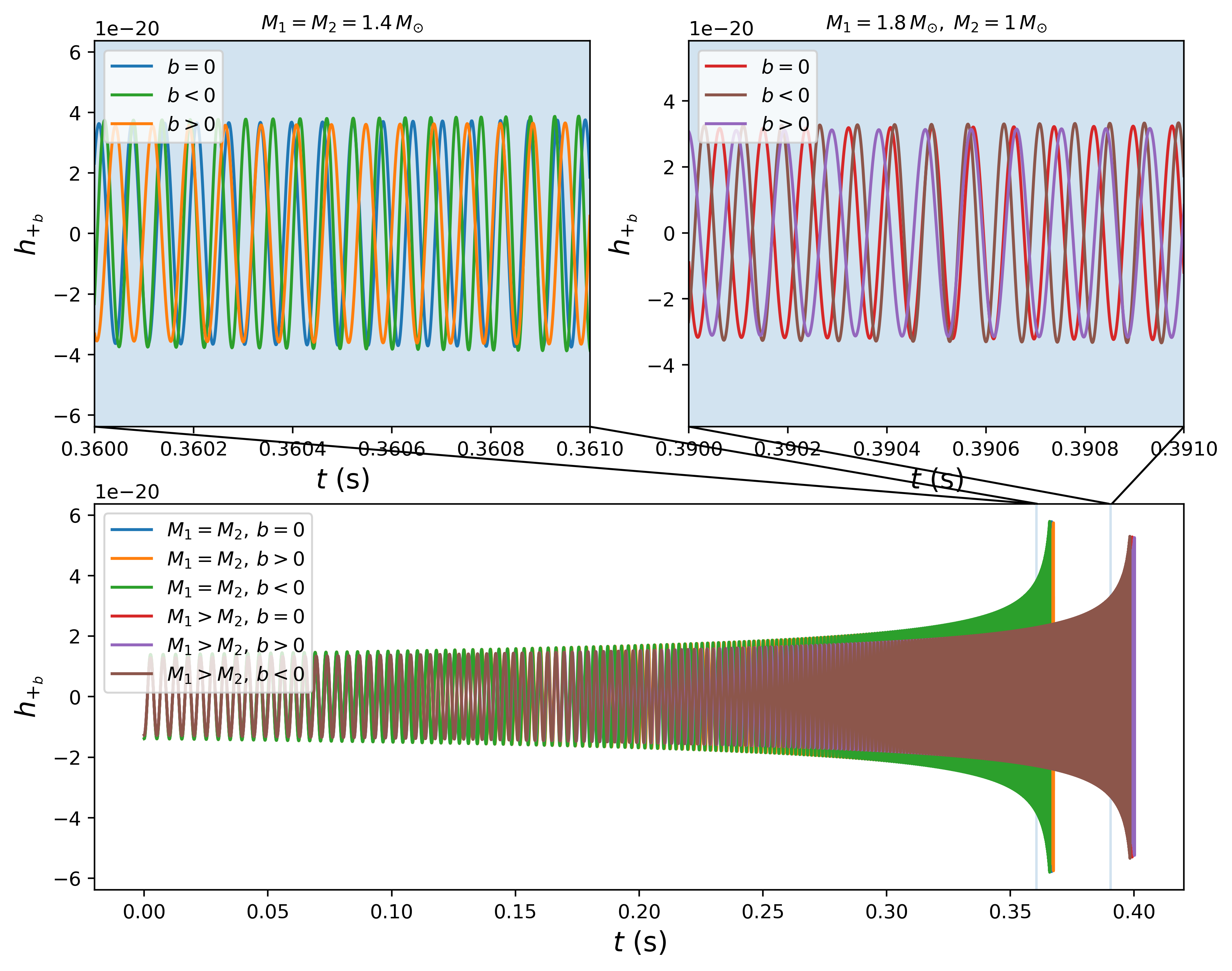}
		\par
	\end{centering}
	\caption{In the figure below, it is shown the $h_+$ polarization of the GW during the inspiral stage for BNS with equal and different masses starting from a initial separation of 100 km. In the top figures, we present a zoom of the final moments, where the effect of the magnetic field is more noticeable.
	}
	\label{fig:hb}
\end{figure}

%%%%%%%%%%%%%%%%%%%%%%%%%%%%%%%%%%%%%%%%%%%%%%%%%%%%%%%%%
In this section we will consider the inspiral stage in circular motion using two BNS systems: one with equal masses of $M_1=M_2=1.4\,M_{\odot}$ ($M=2.8\,M_{\odot}$, $\mu=0.7\,M_{\odot}$) and other with  masses $M_1=1.8\,M_{\odot},\;M_2=1\,M_{\odot}$ ($M=2.8\,M_{\odot}$, $\mu=0.643\,M_{\odot}$) and an initial separation of $r_0=100$ km. We will describe qualitatively the effect of the magnetic field in some relevant variables. We take the radius of both neutron stars as $R=12\,{\rm km}$, and set the minimal radius as $r_{\rm min}=24\,{\rm km}$.  
At this stage, we take a fixed value of the magnetic field $B_1=B_2=8\times10^{16}$\,G and determine the effect of the relative alignment of the dipoles on the inspiral and the consequences in the resulting waveform. 
Using the parameter mentioned before, we calculate the time to reach the minimum radius $\tau_b$, for each configuration with anti-aligned $b<0$ and aligned $b>0$ dipoles. For comparison, we also present the case without magnetic fields $b=0$. The results are shown in Table \ref{tab:tau}. We see that the BNS with equal masses merges before the one with different masses independently of the magnetic fields. In contrast, for both BNS systems, when $b<0$, the time $\tau_b < \tau_ 0$ and when $b> 0$, then $\tau_b>\tau_0$. 
\begin{table}[!ht]
	\centering
	\begin{tabular}{ |c|c|c|c| } 
		\hline
		Magnetic dipole alignment & $b=0$ & $b<0$ & $b>0$ \\
		\hline
		\hline
		Equal masses &$\tau_0=0.36647$ s & $\tau_b=0.36572$ s & $\tau_b=0.36724$ s \\
		Non-equal masses & $\tau_0=0.39905$ s & $\tau_b=0.39815$ s & $\tau_b=0.39996$ s \\
		\hline
	\end{tabular}
	\caption{ Time to reach the minimal radius $r_{\rm min}=24$ km from an initial separation $r_0=100$ km for two relative alignments. The case with $b=0$ correspond to zero magnetic field.}
	\label{tab:tau}
\end{table} 
The separation $r(t)$ is computed from $t=0$ to $\tau_b$ for each case by solving numerically the differential equation (\ref{eq:dr}). In Fig. \ref{fig:insp1} is shown $r(t)$ for cases with equal and unequal masses. 
The relative alignment, as stated above, is given by the sign of $b$.
Additionally, we determine the GW frequency from $\nu_b(t)=\dot{\varphi_b}(r(t))/\pi$ and using Eq. \eqref{eq:dphi}. The resulting frequency as a function of time is plotted in Fig. \ref{fig:nub}. 
The waveforms $h_{+}$ and $h_{\times}$ are obtained from Eq. \eqref{eq:hb} and the numerical integration of Eq. \eqref{eq:dphi} and Eq. \eqref{eq:dr}. In Fig. \ref{fig:hb}, it is plot the polarization $h_{+}$ considering a distance $d=40$ Mpc (the one reported for the detected event GW170817). This distance gives a strain at the minimal radius of $h_b^{\rm (min)}=10^{-20}$, as is show in Fig.~\ref{fig:hb}.
The number of cycles computed from Eq.~(\ref{eq:Ncyc}) are presented in Table \ref{tab:ncyc}. We observe that the BNS with equal masses has less number of cycles ${\cal N}_b$ than the BNS with non-equal masses independently of the magnetic dipole alignment.
We see that regardless the relation between the individual masses, when $b<0$ then ${\cal N}_b < {\cal N}_ 0$ and vise versa, when $b> 0$ then ${\cal N}_b>{\cal N}_0$. 
\begin{table}[!ht]
	\centering
	\begin{tabular}{ |c|c|c|c| } 
		\hline
		Magnetic dipole alignment & $b=0$ & $b<0$ & $b>0$ \\
		\hline 
		\hline
		Equal masses &${\cal N}_0=110.74533$  & ${\cal N}_b=110.46847$  & ${\cal N}_b=111.02361$ \\
		Non-equal masses & ${\cal N}_0=120.58936$ & ${\cal N}_b=120.26116$ & ${\cal N}_b=120.91928$ \\
		\hline
	\end{tabular}
	\caption{Number of cycles in the inspiral stage for both BNS systems, starting to $r_{0}=100$ km. The magnetic dipole alignment corresponds to a field strength $B_1=B_2=8\times 8\times 10^{16} $ G.}
	\label{tab:ncyc}
\end{table}
We have included the cases with equal and different individual masses, which has a more noticeable effect on the collision time in the wave frequency and in the emitted waveform. 

In the next section we shall consider the case with variable magnetic field.

%%%%%%%%%%%%%%%%%%%%%%%%%%%%%%%%%%%%%%%%%%%%
\subsection{Magnetic effect on some inspiral variables}\label{sec:res_bvsX}
\begin{figure}[!ht]
	\begin{centering}
		\includegraphics[scale=.7]{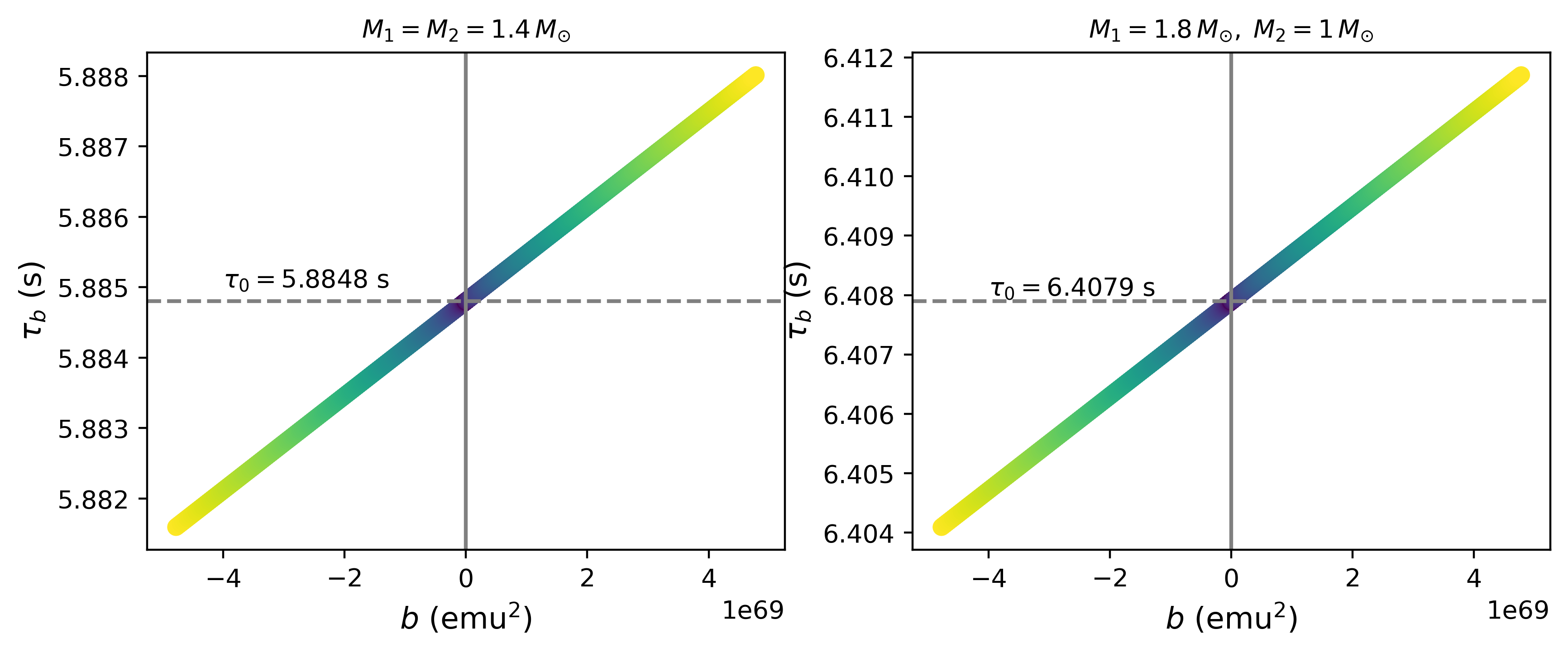}
		\par
	\end{centering}

	\caption{Values of the time at which the ISCO is reached, $\tau_b$,  from an initial separation $r_0=200$ km and a range of values of the magnetic parameter $b$. The code color symbolize the value of the magnetic strength, ranging from purple which corresponds to $B=0$ to yellow corresponding to $8\,\times\,10^{16}$ G.} 
	\label{fig:taub}
\end{figure}

%%%%%%%%%%%%%%%%%%%%%%%%%%%%%%%%%%%%%%%%%%%%
%
%
We proceed further in our analysis.
We now vary the magnitude of the magnetic field and describe the corresponding effect in the inspiral variables: the time to reach the minimal radius, the gravitational luminosity, the strain,  and the frequency of the GWs.
For the analysis in this section, we consider again two BNS systems, one with equal masses of $M_1=M_2=1.4\,M_{\odot}$ and other with  masses  $M_1=1.8\,M_{\odot},\;M_2=1\,M_{\odot}$. 
We consider magnetic fields with strength between $B=10^{12}$ G and $B=8\times 10^{16}$ G, consistent with the values and predictions reported in
\cite{Kaspi:2017fwg,Kaspi:2017zrx}.
These magnitudes of the magnetic field imply that the parameter $b$ is within the interval $-4.77\,\times 10^{69}\; {\rm g\,cm}^5/{\rm s}^2<b<4.77 \times 10^{69}\; {\rm g\,cm}^5/{\rm s}^2.$ 
In Fig.~\ref{fig:taub} the value of $\tau_b$ as a function of $b$ is  plotted for both BNS systems. Note that $\tau_0:=\tau_b(b=0)$. 
As it can be seen in Fig.~\ref{fig:taub}, independently of the magnetic field, the system with equal masses reaches the minimum radius in a shorter time $\tau_b$ than in the case of different masses. However, it is clear in both cases that if $b<0$, then $\tau_b<\tau_0$ and if $b>0$, then $\tau_b>\tau_0$. We can interpret this result as showing that a magnetic configuration with $b>0$ ($b<0$) produces a slight increase (decrease) in the time to reach $r_{min}$. 
This qualitative statement can be quantified through 
the magnetic deviation defined as the ratio:
\begin{equation}\label{eq:dev}
\Delta X = \frac{X_b-X_0}{X_0}, 
\end{equation}
where $X_b$ is a variable that depends on the magnetic parameter $b$ and $X_0=X_b(b=0)$. Following our analysis on $\tau_b$, 
$\Delta \tau = \frac{\tau_b-\tau_0}{\tau_0}$. 
In the cases where $\tau_b>\tau_0$, then $\Delta \tau >0$, this happens for a configuration with $b<0$. In contrast, when $\tau_b<\tau_0$, then $\Delta \tau <0$ which happens for $b>0$. The order of magnitude of $\Delta \tau$ is shown in Table~\ref{dtau_order}, that is the same for BNS with equal and unequal masses. The change in the time to reach the minimum radius can amount to $10^{-4}$ with respect to the change in the time without magnetic influence.

\begin{figure}[!ht]
\begin{centering}
\includegraphics[scale=.7]{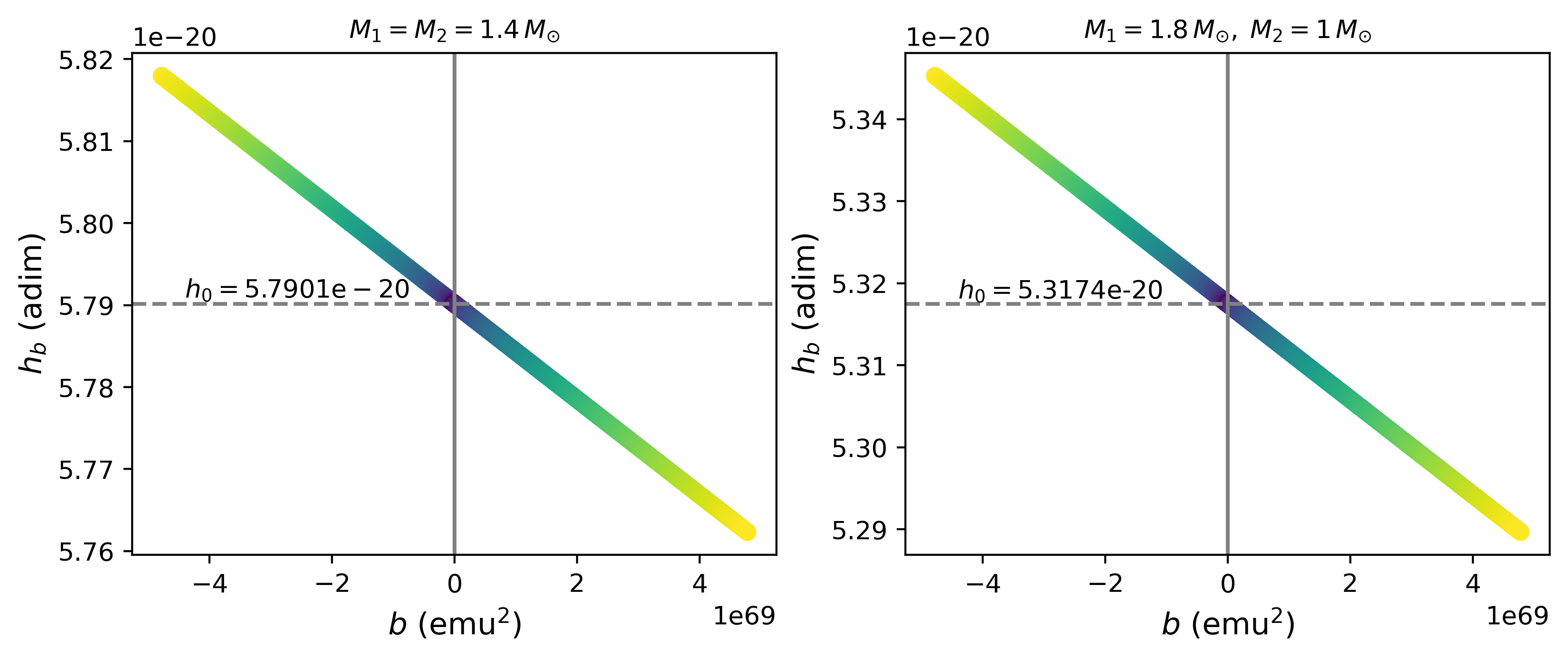} 
\end{centering}
\caption{The strain at $r_{\rm min}$ for both BNS systems. The distance to the source $d$ used to compute the strain is $d=40$ Mpc, similar  to the distance to the source of the signal GW170817. The color code is the same as in figure \ref{fig:taub}.}
\label{fig:h_isco}
\end{figure}

\begin{figure}[!ht]
	\begin{centering}
		\includegraphics[scale=.7]{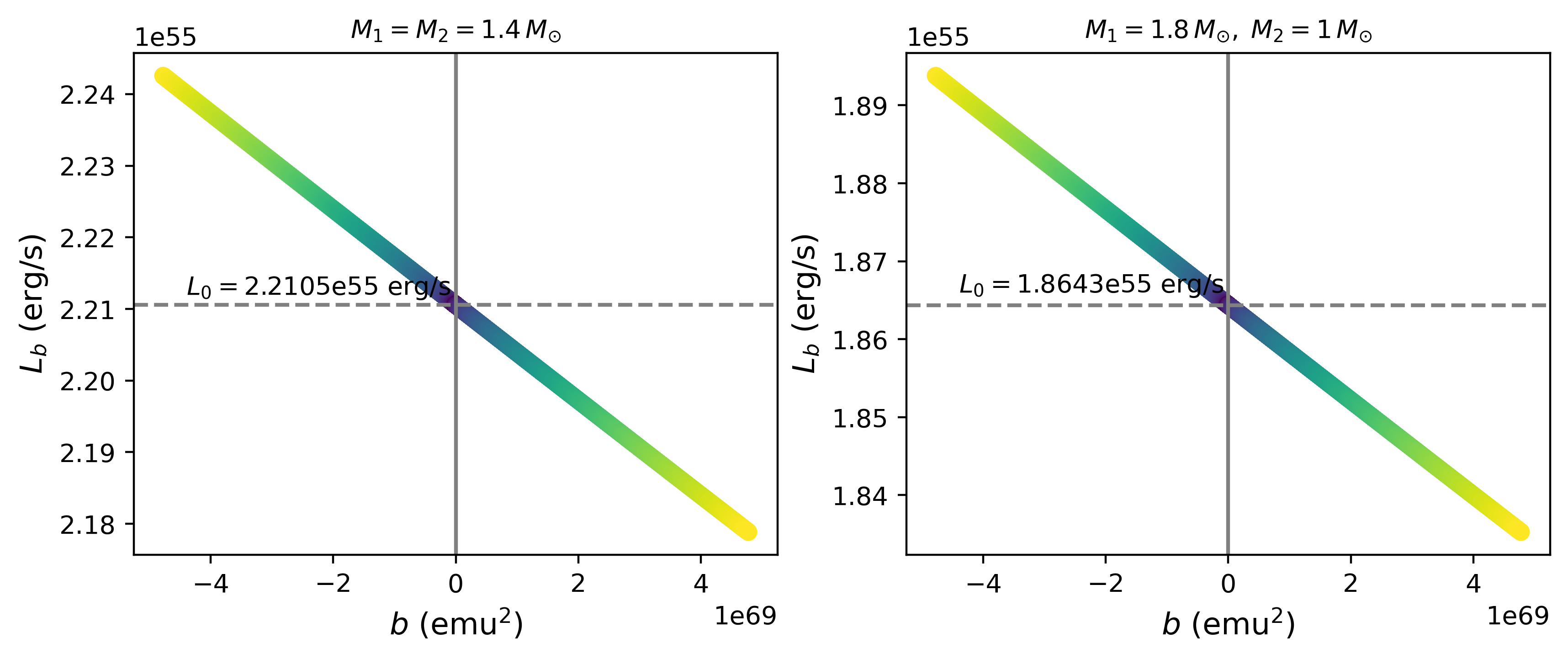}
		\par
	\end{centering}
	\caption{Gravitational luminosity, $L_b(r_{\rm min})$, evaluated at $r_{\rm min}=24.89$ km respect a range of the magnetic parameter $b$ for the BNS with equal masses (left) and unequal masses (right).  
	$L_0$ corresponds to gravitational wave luminosity of the non-magnetized binary. The code color is as in figure \ref{fig:taub}.}
	\label{fig:Lb}
\end{figure}

\begin{figure}[!ht]
	\begin{centering}
		\includegraphics[scale=.7]{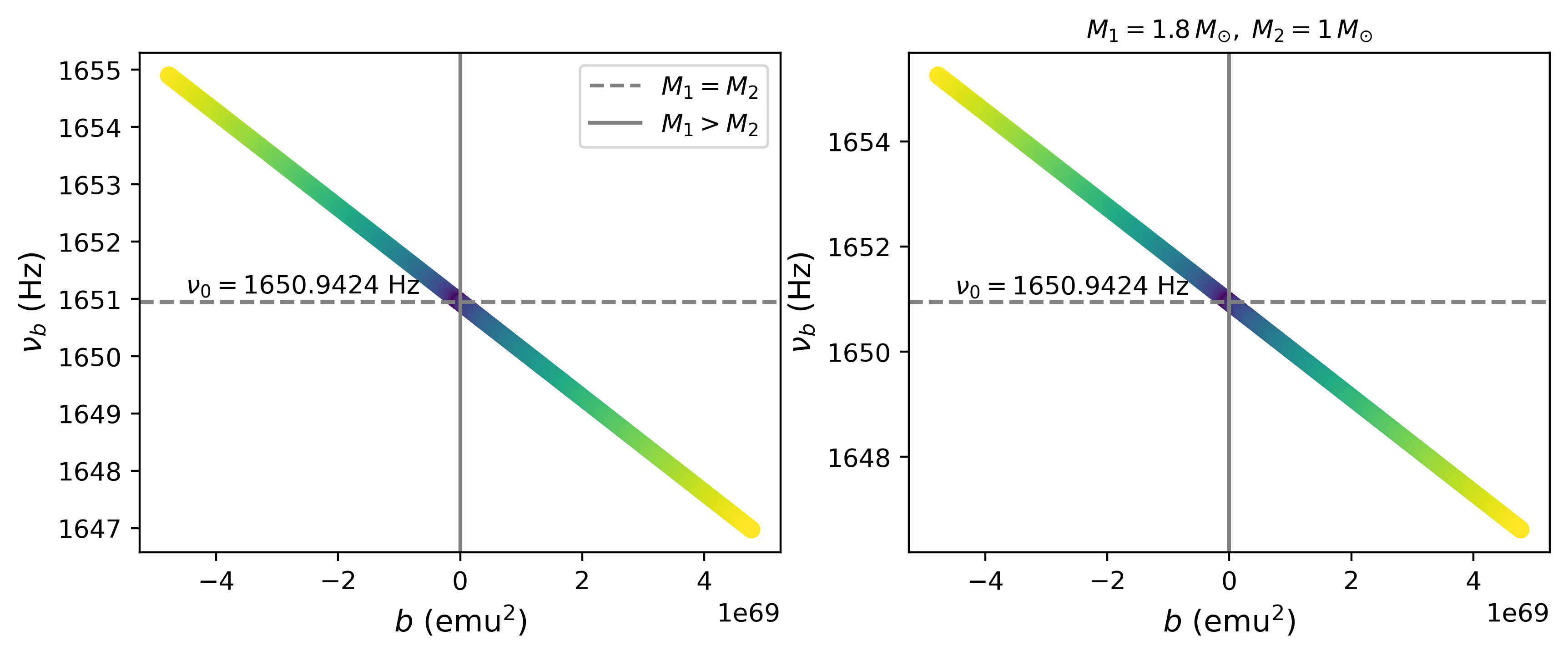}
		\par
	\end{centering}
	\caption{(Left). GW frequency $\nu_b(r)$, evaluated at $r=r_{\rm min}=24.89$ km as a function of the magnetic parameter $b$ for the cases with equal  masses. The frequency $\nu_0(r_{\rm min})$ corresponds to the non-magnetized binary. (Right) The same as Left but for cases with non-equal masses. We use the same code color as in figure \ref{fig:taub}.} 
	\label{fig:fb}
\end{figure}

Let now consider the strain $h_b^{(\rm min)}$ at $r_{\rm min}$ given by Eq. (\ref{eq:hbisco}). If $b<0$, then $h_b^{\rm min}<h_0^{\rm min}$; in contrast if $b>0$, then $h^{\rm min)}_b>h^{\rm min}_0$. In other words, when the magnetic configuration is such that $b<0$,  
the strain of the magnetized system is smaller than the strain of the non-magnetized system at $r_{\rm min}$. The opposite happens when the are such that $b<0$, see Fig. \ref{fig:h_isco}.
\begin{table}[!ht]
	\centering
\begin{tabular}{ |c|c|c|c|c|c|c| } 
\hline
B(G)  & $10^{12}$ & $10^{13}$ & $10^{14}$ & $10^{15}$& $10^{16}$ \\ 
\hline 
\hline
$\sim\,|\Delta \tau|$\, & $10^{-12}$  & $10^{-10}$ & $10^{-8}$ & $10^{-6}$ & $10^{-4}$ \\ 
\hline
$\sim\,|\Delta h(r_{\rm min})|$ & $10^{-11}$  & $10^{-9}$ & $10^{-7}$ & $10^{-5}$ & $10^{-3}$ \\ 
\hline
$\sim |\Delta L(r_{\rm min})|$ & $10^{-10}$  & $10^{-8}$ & $10^{-6}$ & $10^{-4}$ & $10^{-2}$ \\ 
\hline
\end{tabular}
\caption{Amplitude of the deviation on $\tau_b$, $h_b(r_{\rm min})$ and  $L_b(r_{\rm min})$ for some typical values of the magnetic strength $B$ present in NS.}
\label{dtau_order}
\end{table}
The order of magnitude of the magnetic deviation for $\Delta h_b^{\rm min}$ is shown in Table~\ref{dtau_order}.  Note that, in contrast to the deviation on the merger time, the sign of the deviation of the strain at $r_{\rm min}$ is positive when $b<0$, which means that $\Delta h^{\rm min}>0$ when the dipoles are anti-aligned, and  $\Delta h^{\rm min}<0$ when the magnetic dipoles are aligned. 
In this case, the change in the magnitude of the strain can be up to $10^{-3}$.
The luminosity $L_b$ given by Eq. (\ref{lgm}) and the GW frequency $\nu_b$ evaluated at $r_{\rm min}$ are shown in Fig.~\ref{fig:Lb} and Fig. \ref{fig:fb}. 
The GW frequency $\nu_b$ is plotted without the bar color because of the tiny difference between the two BNS systems. 
In Table \ref{dtau_order} are shown the magnetic deviations for $\tau$, $\Delta h$ and $\Delta L$.
For the maximum value of the magnetic strength $B \sim 10^{16}$ G, the merger time
 $\Delta \tau_{b}(r_{\rm min}) \sim 10^{-4}$, the strain $\Delta h (r_{\rm min})\sim 10^{-3}$ and $\Delta L_{b} \sim 10^{-2}$. 
Notice that for $b>0$, 
all deviations, except $\Delta \tau_b$, are negative,  
and the deviation in the non-magnetized cases is most noticeable in the gravitational luminosity $L_{b}$, being the change of order $10^{-2}$. 
The results presented extend all along the entire inspiral stage at any radius, due to the  monotonous behavior of the variables.  
A summary of the qualitative behaviour of the time
$\tau_b$, the strain $h_b$, the gravitational luminosity $L_b$, the GW frequency $\nu_b$, and the number of cycles ${\cal N}_b$
depending on the sign of $b$ is shown in Table \ref{tab:resume}. 
\begin{table}[!ht]
	\centering
\begin{tabular}{ |c|c|c|c| } 
\hline
Variable & $b=0$ & $b<0$ & $b>0$ \\
\hline
\hline
$\tau_b$ & $\tau_0$ & $\tau_b<\tau_0$ & $\tau_b>\tau_0$ \\ 
\hline
$h_b$ & $h_0$ & $h_b>h_0$ & $h_b<h_0$ \\ 
\hline
$L_b$ & $L_0$ & $L_b>L_0$ & $L_b<L_0$ \\ 
\hline
$\nu_b$ & $\nu_0$ & $\nu_b>\nu_0$ & $\nu_b<\nu_0$ \\ 
\hline
${\cal N}_b$ & ${\cal N}_0$ & ${\cal N}_b<{\cal N}_0$ & ${\cal N}_b>{\cal N}_0$ \label{tableII} \\ 
\hline 
\end{tabular}
\caption{
Relative behaviour of some astrophysical variables; the time to reach $r_{\rm min}$, $\tau_b$, the strain, $h_b$, the gravitational luminosity, $L_b$, the GW frequency $\nu_b$ and the number of cycles ${\cal N}_b$,
between magnetized and non magnetized cases.
}
\label{tab:resume}
\end{table}

%------------------------------------------------------
%%%%%%%%%%%%%%%%%%%%%%%%%%%%%%%%%%%%%%%%%%%%%%%%%%%%%%%%%%
\subsection{Mass estimation: Another application.} 
\label{sub:mass_estimation}
%%%%%%%%%%%%%%%%%%%%%%%%%%%%%%%%%%%%%%%%%%%%%%%%%%%%%%%%%%
%
The scenario presented in this part is slightly different to the one presented previously. Here we show how  that
the uncertainties on the determination on the mass of a astrophysical system may be used to impose bounds on the maximum value of the magnetic field presented in a BNS system.
We will consider for such purpose the  GW170817 signal
due its importance as
the first evidence of the collision of two NS and its relevance in ongoing astrophysical developments. 
Furthermore, some of the assumptions used in the present work (such as zero eccentricity and non spinning progenitors) are consistent with the ones reported in \cite{LIGOScientific:2017vwq}.
We proceed as follows: 
LVC reports for GW170817 a total mass of $M=2.74_{-0.01}^{+0.04}\,M_{\odot}$ for the progenitor BNS \cite{LIGOScientific:2017vwq}. Additionally, the statistical and systematic errors imply two limit values for the total mass $$M_{\rm min}=2.73\,M_{\odot}\qquad  {\rm and}  \qquad M_{\rm max}=2.78\,M_{\odot} \ .$$
Considering the definition of the function $f_M\equiv M/M_0$, and taking the average mass as $M_0=2.74\,M_{\odot}$ and $M_{\rm min} $($M_{\rm max}$) as the minimum (maximum) possible total mass $M$. 
Then 
\begin{equation} \label{fmin}
 f_M^{({\rm min})}\equiv \frac{M_{\rm min}}{M_0}=\frac{2.73}{2.74}=0.99635 \ ,   
\end{equation}
and
\begin{equation}\label{fmax}
f_M^{({\rm max})}\equiv \frac{M_{\rm max}}{M_0}=\frac{2.78}{2.74}=1.0145 \ .    
\end{equation}
This provides a domain in $x$ for the function $f_M$, $$0.99635\leq f_M \leq 1.0145.$$ which is still consistent with the data.
Using the fact that $f_M$ is  monotonous on $x$ in a small vicinity of $x=0$, (this can be seen from Eq. (\ref{eq:cM})), it can be inverted to get the maximum and minimum values of $x$, as
$x_{\rm min}=-0.00225$ {\rm and} $x_{\rm max}=0.00058$.
\begin{figure}[ht]
	\begin{centering}
		\includegraphics[scale=.7]{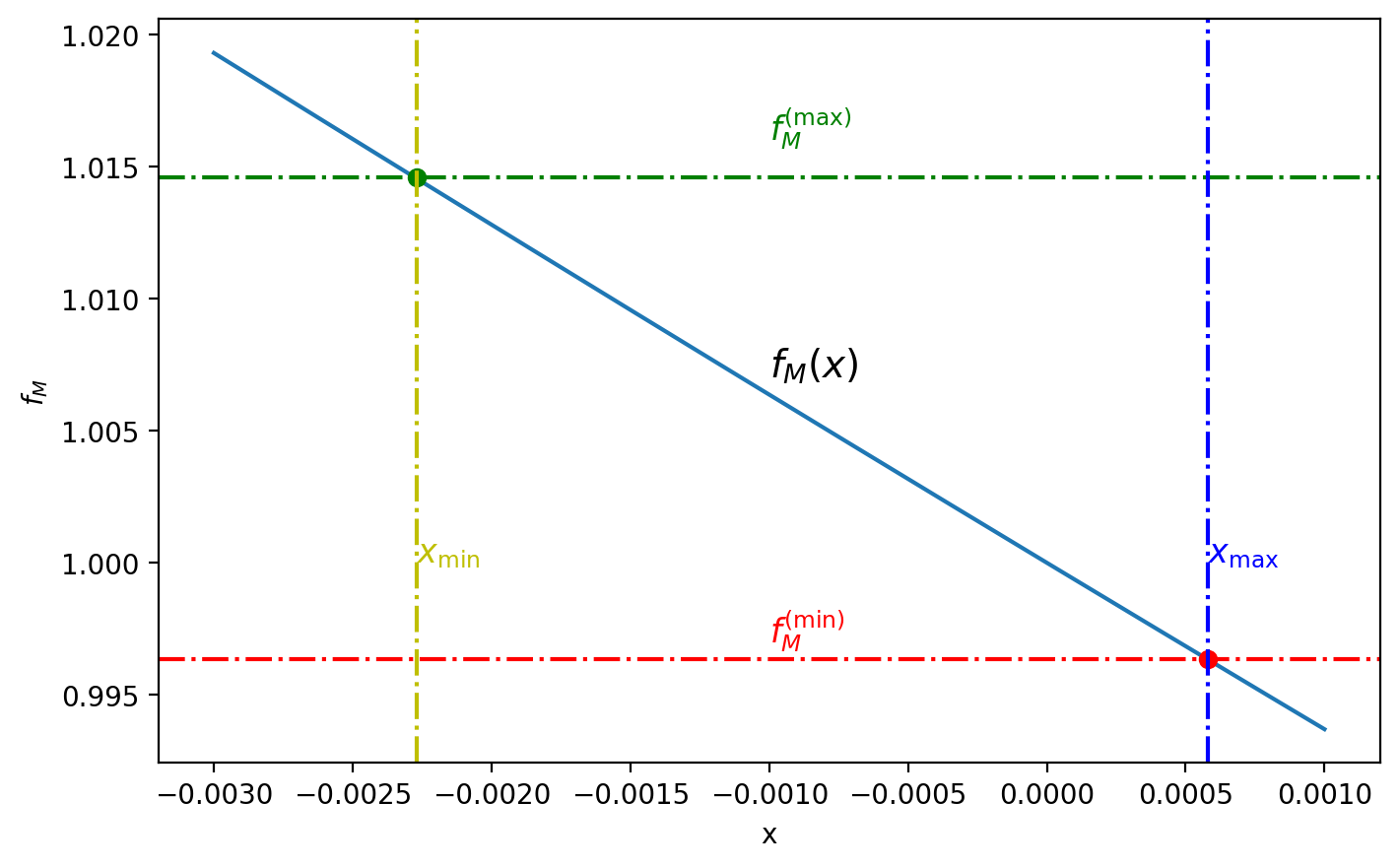}
		\par
	\end{centering}
	\caption{
	Function $f_M(x)$ with $L$ and $Q$ of estimated from the source signal GW170817. Horizontal lines represent
	$f_M^{({\rm max})}$ and $f_M^{({\rm min})}$ as defined in Eq. \eqref{fmax} and Eq.\eqref{fmin} respectively. Vertical lines are used to better identify the values of $x_{\rm max}$ and $x_{\rm min}$.
	The intersection of the three lines determine
	the vales of $f_M$ that allow masses consistent with GW170917.
	}
	\label{fig:fMgw}
\end{figure}
Next, from the definition of the dimensionless variable $x$, Eq.~(\ref{eq:x})
we obtain that $b=Lr^3 x/Q.$ Using the values consistent with GW170817 for the luminosity and the logarithmic change in the period;
$L=1.8423\times10^{55}$ ergs/s and 
$Q=-272.4447$ Hz, we obtain the following range for
the magnetic parameter 
\begin{eqnarray}
b_{\rm min}&=&-9.69\times10^{71}x_{\rm max} {\rm emu}^2 =-5.623\times10^{68}{\rm emu}^2, \nonumber \\
b_{\rm max}&=&-9.69\times10^{71}x_{\rm min} {\rm emu}^2=2.18\times10^{69}{\rm emu}^2.
\end{eqnarray}
Note that the sign of $b$ is opposite to the sign of $x$ such that that $b_{\rm min}<0$ corresponds to a $x_{\rm max}>0$ and vice versa. 
From Eq.~(\ref{eq:b}), we can finally, estimate the strength of the magnetic fields associated with the maximum and minimum values of the mass. 
In terms of the magnetic field of the dipoles,
the value of $b_{\rm min}$ requires anti-aligned dipoles with magnetic fields of strength $B=2.74\times10^{16}$ G, the maximum theorized for NS, and $b_{\rm max}$ requires aligned dipoles with magnetic fields of  $B=5.41\times10^{16}$ G. 
In summary, the highest value of the mass for GW170817 can be associated to the presence of aligned magnetic fields. Whereas the lowest possible total mass can be associated to anti-aligned magnetic fields of the same strength. 
It can be infer thus, the uncertainties in the LVC mass determination, allow for the presence of magnetic fields as large as $B=2.74\times10^{16}$. It is important that, as the precision in the measurements is increased, the strength of the magnetic field involved in the BNS will be more accurately determined.

%%%%%%%%%%%%%%%%%%%%%%%%
\section{CONCLUSIONS} 
\label{sec:conclusions}
%%%%%%%%%%%%%%%%%%%%%%%%

Neutron stars are amongst the astrophysical objects with the strongest magnetic fields in the universe. It is thus expected magnetic fields play a major role in their dynamics. In binary systems the GW emission is slightly affected due to the presence of strong magnetic fields as has been described in several works \cite{Woods:2004kb, Ioka:2000yb}.
In this work we presented a simple, but still useful, model of a BNS system that incorporates a magnetic field in the dynamics of the binary. 
Our approach is based on the quadrupolar formalism to calculate the GW emitted during the inspiral phase of a magnetized BNS system. We use the newtonian description of gravity and describe the magnetic field of each star as a magnetic dipole. Furthermore, since the GW emission tends to circularize the orbit of the binary we focus only on the circular case.
Assuming that the contribution of the individual spins to the total angular momentum of the binary is very small compared to the orbital angular momentum, that is the dominant contribution to the total angular momentum is the orbital one, the equations that describe the gravitational and electromagnetic interaction can be cast in a very simple form.
In particular, it has been argued that in the (magnetic) dipole approximation for the NS, the individual magnetic moments remain aligned to the orbital angular momentum. Under this consideration we have shown that the dynamics of the binary can be described in terms of an effective one-body problem.
By studying the effective potential of the equivalent one body problem, we showed there is a critical value for the magnetic field strength below which no bound orbits exist. 
This critical value arises only when the magnetic moments are anti-aligned. 
That is, 
a bounded system does not exist if the magnetic field is strong enough, because the magnetic repulsion is strong enough to overcome the gravitational attraction.
In our analysis we determined the effect of the magnetic field in some  astrophysical relevant quantities of the binary such as
GW luminosity, the logarithmic rate of change of the orbital period, the time to reach the minimum radius and total mass. As expected the results presented here reduce to the circular binary problem described for instance in \cite{Maggiore:2007ulw} in the absence of magnetic fields. 
As an application of our model, we showed that
for BNS systems with total mass $M= 2.8\,M_{\odot}$, and a strength of the magnetic fields of $B\simeq10^{16}$ G the ratio between the magnetic and gravitational potential $U_m/U_g$ is of the order of $\sim 10^{-4}$ when the stars are in the verge on collision.
We have also found that deviations in the frequency and strain of the GW with respect to the non magnetic case, are of the order of $\sim 10^{-4}$ and  the deviations in the luminosity may be as large as $\sim 10^{-2}$ 
with respect to the non magnetized system.
Furthermore we apply the model to two possible astrophysical scenarios. 
(i) Considering that the gravitational luminosity, and the logarithmic rate of change of the orbital period can be extracted from observational data, our model allows us to determinate the effect of the magnetic field in the determination of the individual masses on the binary. We showed that the total and the reduced masses, are over or sub estimated with respect to systems where there is no magnetic interaction.  Additionally, if the magnetic moments are aligned, the total mass is sub-estimated whereas if the magnetic moments are anti-aligned the total mass is over-estimated. The opposite happens with the reduced mass. In fact, a strength of $B\sim 10^{16}$ G in the magnetic field may cause an over-estimation or sub-estimation up to $2\%$ of the total mass. Although the percentage seems small, the deviation falls within the uncertainty ranges of the LVC detectors \cite{LIGOScientific:2021djp}.
(ii)
We use the event GW170817 as a test case in our model and showed that the uncertainly in the determination on the mass of the progenitors reported in the literature imposes naturally a range for the possible values of the magnetic field strength.
In summary, the maximum and minimum possible values of the total mass associated to GW170817 allowed us to impose a maximum value for the magnetic strength of $B\sim10^{16}$ G.

As estimated in Sec. \ref{sec:results} magnetic field effects are less likely to be detectable by the current observatories unless the magnetic field strength is as high as $B=10^{17}$ G.
Since future improvements in GW observations will allow to determine the effect of magnetic fields on the properties of the binary accurately, studies like the one presented here will be of the upmost importance. Based on our results, templates of the waveform can be generated and proceed to carry out the corresponding search making use of the data released from the LIGO collaboration. 

%%%%%%%%%%%%%%%%%%%%%%%%%%%%%%%%%%%%%
\begin{acknowledgments}
%%%%%%%%%%%%%%%%%%%%%%%%%%%%%%%%%%%%%
We are grateful to Francisco S. Guzm\'an, Celia Escamilla, Diego L\'opez, and Dany Page for fruitful comments during the elaboration of the present manuscript. This work was supported in part by the CONACYT Network Projects No. 376127 
``Sombras, lentes y ondas gravitatorias generadas por objetos compactos astrof\'isicos'', No. 304001 ``Estudio de campos escalares con aplicaciones en cosmolog\'ia y astrof\'isica'' and No. 1406300 ``Explorando los confines de las teor\'ias
relativistas de la gravitaci\'on y sus consecuencias'', as well as by DGAPA-UNAM through grants 
IN110218, IA103616, IN105920. We also acknowledge support from the European Union s Horizon 2020 research and innovation (RISE)
program H2020-MSCA-RISE-2017 Grant
No. FunFiCO-777740. ML acknowledges support from CONACYT graduate  grants  program.
\end{acknowledgments}

%%%%%%%%%%%%%%%%%%%%%%%%%%%%%%%%%%%%%%%%%%%%%%%%%%%%%%%
\appendix
\section{A comment on the rotation of NS}
\label{ap:rot_NS}
%%%%%%%%%%%%%%%%%%%%%%%%%%%%%%%%%%%%%%%%%%%%%%%%%%%%%%%

The pulsar paradigm states that pulsars are rotating NS, which can successfully explain the astrophysical observations, as the pulsar spin period $P_s$ and its extreme magnetic fields. 

On the one hand, in the Newtonian regime, if the spins ${\bf S}_i$ ($i=1,2$) are considered, the orbital angular momentum ${\bf L}$ is not the quantity conserved, but the total angular momentum $\mathbf{J}$ given by $\mathbf{J}={\bf S}_1+{\bf S}_2+{\bf L}$. Spin ${\bf S}_i$ is related to the angular frequency ${\bf \Omega}_i$ thought ${\bf S}_i = I_i{\bf \Omega}_i$. The rotation provides a kinetic rotational energy given by $E_{{\rm rot}_i}= I_i \Omega_i^2 /2$, where $\Omega_i=|{\bf \Omega}_i|=2\pi/P_{s_i}$, $P_{s_i}$ is the spin period and $I_i$ is its moment of inertia\footnote{For NS $I_i=a_i(x) M_i R_i^2$, where $x$ is a dimensionless compactness parameter $a_i$ that depends of the EoS chosen, $0\leq a\leq1$.}. Then, total rotational energy $E_{{\rm rot}}=E_{{\rm rot}_1}+E_{{\rm rot}_2}$ must be added to the Lagrangian function given in Eq. (\ref{eq:lag}) of each star. If one not restring the direction of the spins, the systems has an least seven Euler-Lagrange equations: one for $r$, six for both ${\bf S}_i$. 

On the other hand, in inspiral stage, spins introduces a spin-orbit and spin-spin coupling in the BNS dynamic and waveform. These complex relations causes that if the spins are not aligned to the orbital angular momentum, then the spins and the orbital plane of the binary to precess \cite{Apostolatos:1994ha}, i.e., both the spins ${\bf S}_i$ and the orbital angular momentum $\mathbf{L}$ precesses about the total angular momentum. 

However, in recent works had found that in inspiral stage, precession effects are little and approximately decouple in the BNS dynamics. Numerical simulations have shown that the direction of the angular momentum is conserved during inspiral stage \cite{Capozziello:2011fs,Hannam:2013oca}. In fact, the state-of-the art in the effective-one-body BNS waveforms models are constructed twisted up non-precessing binary waveforms with approximate expressions for the precessional motion \cite{Hannam:2013oca}. Also, it has been found that the temporal evolution of the total angular momentum is principally due to the loss of orbital angular moment, $\dot{\mathbf{L}}$ (where dot means derivative on time). It occurs due to the fact that the spins changes at a much greater scale of time, so that the objects cannot be spun up substantially during the seconds of the inspiral phase. Hence, in general in the waveform models is assumed that $\dot{\mathbf{S}}_i\simeq0$, and $\dot{\mathbf{J}}=\dot{\mathbf{L}}$. 

Otherwise, if each NS rotates, then its  magnetic moment ${\bf m}_i$ depends of time and according to classical electrodynamics, the NS emits electromagnetic waves. The Larmor formula states that the electromagnetic luminosity $L_{{\rm EM}_i}$ is described by $L_{{\rm EM}_i}=(2/3)|\ddot{{\bf m}}_i|^2/c^3$. For a perfect magnetic dipole,  ${\bf m}_i(t) = (1/2)B_i R_i^3 \sin\chi_i \exp^{\rm i \Omega_i t} \hat{{\bf z}}$, where  ${\bf \Omega}_i=\Omega_i \,\hat{{\bf z}}$ and $\chi_i$ the angle between ${\bf S}_i$ and ${\bf m}_i$, then the electromagnetic luminosity is \cite{Zhang:2000wx} $$L_{{\rm EM}_i}=\frac{B_i^2 R_i^6 \Omega_i ^4\, \sin^2 \chi_i}{6c^3}.$$ 

Moreover, if each NS is a nonsymmetric object and the rotation axis is not a symmetry axis, the rotational energy also can be released thought GWs. This other gravitational luminosity is described by \cite{Shapiro:1983du} $$L_{GW_i}=\frac{32GI_i^2 \epsilon_i^2 \Omega_i^6}{5c^5},$$ where $\epsilon$  is the ellipticity of NS defined as  $\epsilon \equiv (I_1-I_2)/I_3$, $I_n$ are the principal moments of inertia ($n=1,2,3$). In fact, the balance of rotational energy says that $$-\dot{E}_{rot_i}=L_{{\rm EM}_i}+L_{{\rm GW}_i},$$ where $L_{{\rm EM}_i}$ is a classical electromagnetic luminosity and $L_{{\rm GW}_i}$ is the gravitational luminosity, which gives a index law for the angular frequency $\Omega_i$. 

In our model, we ignore these radiative phenomenon, but if we incorporate these, the inspiral balance of energy equation should changes to $$\dot{E_T}=L_{{\rm GW}_b}+L_{{\rm EM}_1}+L_{{\rm GW}_1}+L_{{\rm EM}_2}+L_{{\rm GW}_2},$$ where $E_T=E_b+E_{{\rm rot}_1}+E_{{\rm rot}_2}$, $L_{{\rm GW}_b}$ is given by the Eq. (\ref{eq:L}) and $E_b$ by the Eq. (\ref{eq:Ec}). However, for simplicity one can suppose that the spins are aligned with the orbital angular momentum ${\bf L}$. Otherwise, one could construct the general case added approximate expression for the precessional motion. 

From our assumptions, also the magnetic moments are aligned with the orbital angular momentum, so the spins and magnetic moments are aligned: ${\bf m}_i || {\bf S}_i$. It implies not electromagnetic luminosity because $\chi_i=0$. Regarding the individual gravitational luminosity, we can say that it is emitted in another time scale, so in general, one can decouple the balance equations. Thus, the orbital dynamic of the sec. \ref{sec:setup} and the GWs estimates of the sec. \ref{sec:grav_emission} remains valid. 

%%%%%%%%%%%%%%%%%%%%%%
%%%   REFERENCES   %%%
%%%%%%%%%%%%%%%%%%%%%%

\bibliographystyle{apsrev}
\bibliography{aipsamp}

\providecommand{\noopsort}[1]{}\providecommand{\singleletter}[1]{#1}%
\begin{thebibliography}{54}
\expandafter\ifx\csname natexlab\endcsname\relax\def\natexlab#1{#1}\fi
\expandafter\ifx\csname bibnamefont\endcsname\relax
  \def\bibnamefont#1{#1}\fi
\expandafter\ifx\csname bibfnamefont\endcsname\relax
  \def\bibfnamefont#1{#1}\fi
\expandafter\ifx\csname citenamefont\endcsname\relax
  \def\citenamefont#1{#1}\fi
\expandafter\ifx\csname url\endcsname\relax
  \def\url#1{\texttt{#1}}\fi
\expandafter\ifx\csname urlprefix\endcsname\relax\def\urlprefix{URL }\fi
\providecommand{\bibinfo}[2]{#2}
\providecommand{\eprint}[2][]{\url{#2}}

\bibitem[{\citenamefont{Abbott et~al.}(2016)}]{LIGOScientific:2016aoc}
\bibinfo{author}{\bibfnamefont{B.~P.} \bibnamefont{Abbott}}
  \bibnamefont{et~al.} (\bibinfo{collaboration}{LIGO Scientific, Virgo}),
  \bibinfo{journal}{Phys. Rev. Lett.} \textbf{\bibinfo{volume}{116}},
  \bibinfo{pages}{061102} (\bibinfo{year}{2016}), \eprint{1602.03837}.

\bibitem[{\citenamefont{Abbott et~al.}(2009)}]{LIGOScientific:2007fwp}
\bibinfo{author}{\bibfnamefont{B.~P.} \bibnamefont{Abbott}}
  \bibnamefont{et~al.} (\bibinfo{collaboration}{LIGO Scientific}),
  \bibinfo{journal}{Rept. Prog. Phys.} \textbf{\bibinfo{volume}{72}},
  \bibinfo{pages}{076901} (\bibinfo{year}{2009}), \eprint{0711.3041}.

\bibitem[{\citenamefont{Aasi et~al.}(2015)}]{LIGOScientific:2014pky}
\bibinfo{author}{\bibfnamefont{J.}~\bibnamefont{Aasi}} \bibnamefont{et~al.}
  (\bibinfo{collaboration}{LIGO Scientific}), \bibinfo{journal}{Class. Quant.
  Grav.} \textbf{\bibinfo{volume}{32}}, \bibinfo{pages}{074001}
  (\bibinfo{year}{2015}), \eprint{1411.4547}.

\bibitem[{\citenamefont{Abbott
  et~al.}(2017{\natexlab{a}})}]{LIGOScientific:2017vwq}
\bibinfo{author}{\bibfnamefont{B.~P.} \bibnamefont{Abbott}}
  \bibnamefont{et~al.} (\bibinfo{collaboration}{LIGO Scientific, Virgo}),
  \bibinfo{journal}{Phys. Rev. Lett.} \textbf{\bibinfo{volume}{119}},
  \bibinfo{pages}{161101} (\bibinfo{year}{2017}{\natexlab{a}}),
  \eprint{1710.05832}.

\bibitem[{\citenamefont{Abbott
  et~al.}(2017{\natexlab{b}})}]{LIGOScientific:2017ync}
\bibinfo{author}{\bibfnamefont{B.~P.} \bibnamefont{Abbott}}
  \bibnamefont{et~al.} (\bibinfo{collaboration}{LIGO Scientific, Virgo, Fermi
  GBM, INTEGRAL, IceCube, AstroSat Cadmium Zinc Telluride Imager Team, IPN,
  Insight-Hxmt, ANTARES, Swift, AGILE Team, 1M2H Team, Dark Energy Camera
  GW-EM, DES, DLT40, GRAWITA, Fermi-LAT, ATCA, ASKAP, Las Cumbres Observatory
  Group, OzGrav, DWF (Deeper Wider Faster Program), AST3, CAASTRO, VINROUGE,
  MASTER, J-GEM, GROWTH, JAGWAR, CaltechNRAO, TTU-NRAO, NuSTAR, Pan-STARRS,
  MAXI Team, TZAC Consortium, KU, Nordic Optical Telescope, ePESSTO, GROND,
  Texas Tech University, SALT Group, TOROS, BOOTES, MWA, CALET, IKI-GW
  Follow-up, H.E.S.S., LOFAR, LWA, HAWC, Pierre Auger, ALMA, Euro VLBI Team, Pi
  of Sky, Chandra Team at McGill University, DFN, ATLAS Telescopes, High Time
  Resolution Universe Survey, RIMAS, RATIR, SKA South Africa/MeerKAT}),
  \bibinfo{journal}{Astrophys. J. Lett.} \textbf{\bibinfo{volume}{848}},
  \bibinfo{pages}{L12} (\bibinfo{year}{2017}{\natexlab{b}}),
  \eprint{1710.05833}.

\bibitem[{\citenamefont{Abbott et~al.}(2020)}]{LIGOScientific:2020aai}
\bibinfo{author}{\bibfnamefont{B.~P.} \bibnamefont{Abbott}}
  \bibnamefont{et~al.} (\bibinfo{collaboration}{LIGO Scientific, Virgo}),
  \bibinfo{journal}{Astrophys. J. Lett.} \textbf{\bibinfo{volume}{892}},
  \bibinfo{pages}{L3} (\bibinfo{year}{2020}), \eprint{2001.01761}.

\bibitem[{\citenamefont{Sinha et~al.}(2013)\citenamefont{Sinha, Mukhopadhyay,
  and Sedrakian}}]{Sinha:2010fm}
\bibinfo{author}{\bibfnamefont{M.}~\bibnamefont{Sinha}},
  \bibinfo{author}{\bibfnamefont{B.}~\bibnamefont{Mukhopadhyay}},
  \bibnamefont{and}
  \bibinfo{author}{\bibfnamefont{A.}~\bibnamefont{Sedrakian}},
  \bibinfo{journal}{Nucl. Phys. A} \textbf{\bibinfo{volume}{898}},
  \bibinfo{pages}{43} (\bibinfo{year}{2013}), \eprint{1005.4995}.

\bibitem[{\citenamefont{{Arzoumanian} et~al.}(1994)\citenamefont{{Arzoumanian},
  {Nice}, {Taylor}, and {Thorsett}}}]{1994ApJ...422..671A}
\bibinfo{author}{\bibfnamefont{Z.}~\bibnamefont{{Arzoumanian}}},
  \bibinfo{author}{\bibfnamefont{D.~J.} \bibnamefont{{Nice}}},
  \bibinfo{author}{\bibfnamefont{J.~H.} \bibnamefont{{Taylor}}},
  \bibnamefont{and} \bibinfo{author}{\bibfnamefont{S.~E.}
  \bibnamefont{{Thorsett}}}, \bibinfo{journal}{\apj}
  \textbf{\bibinfo{volume}{422}}, \bibinfo{pages}{671} (\bibinfo{year}{1994}).

\bibitem[{\citenamefont{Vigan\`o}(2013)}]{Vigano:2013uia}
\bibinfo{author}{\bibfnamefont{D.}~\bibnamefont{Vigan\`o}},
  \bibinfo{type}{Other thesis}, \bibinfo{school}{Universitat d'Alacant}
  (\bibinfo{year}{2013}), \eprint{1310.1243}.

\bibitem[{\citenamefont{Kaspi}(2017)}]{Kaspi:2017zrx}
\bibinfo{author}{\bibfnamefont{V.~M.} \bibnamefont{Kaspi}},
  \bibinfo{journal}{IAU Symp.} \textbf{\bibinfo{volume}{337}},
  \bibinfo{pages}{3} (\bibinfo{year}{2017}), \eprint{1806.03697}.

\bibitem[{\citenamefont{Woods and Thompson}(2004)}]{Woods:2004kb}
\bibinfo{author}{\bibfnamefont{P.~M.} \bibnamefont{Woods}} \bibnamefont{and}
  \bibinfo{author}{\bibfnamefont{C.}~\bibnamefont{Thompson}}
  (\bibinfo{year}{2004}), \eprint{astro-ph/0406133}.

\bibitem[{\citenamefont{Duncan and Thompson}(1992)}]{Duncan:1992hi}
\bibinfo{author}{\bibfnamefont{R.~C.} \bibnamefont{Duncan}} \bibnamefont{and}
  \bibinfo{author}{\bibfnamefont{C.}~\bibnamefont{Thompson}},
  \bibinfo{journal}{Astrophys. J. Lett.} \textbf{\bibinfo{volume}{392}},
  \bibinfo{pages}{L9} (\bibinfo{year}{1992}).

\bibitem[{\citenamefont{Kaspi and Beloborodov}(2017)}]{Kaspi:2017fwg}
\bibinfo{author}{\bibfnamefont{V.~M.} \bibnamefont{Kaspi}} \bibnamefont{and}
  \bibinfo{author}{\bibfnamefont{A.}~\bibnamefont{Beloborodov}},
  \bibinfo{journal}{Ann. Rev. Astron. Astrophys.}
  \textbf{\bibinfo{volume}{55}}, \bibinfo{pages}{261} (\bibinfo{year}{2017}),
  \eprint{1703.00068}.

\bibitem[{\citenamefont{Ioka and Taniguchi}(2000)}]{Ioka:2000yb}
\bibinfo{author}{\bibfnamefont{K.}~\bibnamefont{Ioka}} \bibnamefont{and}
  \bibinfo{author}{\bibfnamefont{K.}~\bibnamefont{Taniguchi}},
  \bibinfo{journal}{Astrophys. J.} \textbf{\bibinfo{volume}{537}},
  \bibinfo{pages}{327} (\bibinfo{year}{2000}), \eprint{astro-ph/0001218}.

\bibitem[{\citenamefont{Price and Rosswog}(2006)}]{Price:2006fi}
\bibinfo{author}{\bibfnamefont{D.}~\bibnamefont{Price}} \bibnamefont{and}
  \bibinfo{author}{\bibfnamefont{S.}~\bibnamefont{Rosswog}},
  \bibinfo{journal}{Science} \textbf{\bibinfo{volume}{312}},
  \bibinfo{pages}{719} (\bibinfo{year}{2006}), \eprint{astro-ph/0603845}.

\bibitem[{\citenamefont{Troja et~al.}(2010)\citenamefont{Troja, Rosswog, and
  Gehrels}}]{Troja:2010zm}
\bibinfo{author}{\bibfnamefont{E.}~\bibnamefont{Troja}},
  \bibinfo{author}{\bibfnamefont{S.}~\bibnamefont{Rosswog}}, \bibnamefont{and}
  \bibinfo{author}{\bibfnamefont{N.}~\bibnamefont{Gehrels}},
  \bibinfo{journal}{Astrophys. J.} \textbf{\bibinfo{volume}{723}},
  \bibinfo{pages}{1711} (\bibinfo{year}{2010}), \eprint{1009.1385}.

\bibitem[{\citenamefont{Lipunov and Panchenko}(1996)}]{Lipunov:1996wf}
\bibinfo{author}{\bibfnamefont{V.~M.} \bibnamefont{Lipunov}} \bibnamefont{and}
  \bibinfo{author}{\bibfnamefont{I.~E.} \bibnamefont{Panchenko}},
  \bibinfo{journal}{Astron. Astrophys.} \textbf{\bibinfo{volume}{312}},
  \bibinfo{pages}{937} (\bibinfo{year}{1996}), \eprint{astro-ph/9608155}.

\bibitem[{\citenamefont{Hansen and Lyutikov}(2001)}]{Hansen:2000am}
\bibinfo{author}{\bibfnamefont{B.~M.~S.} \bibnamefont{Hansen}}
  \bibnamefont{and} \bibinfo{author}{\bibfnamefont{M.}~\bibnamefont{Lyutikov}},
  \bibinfo{journal}{Mon. Not. Roy. Astron. Soc.}
  \textbf{\bibinfo{volume}{322}}, \bibinfo{pages}{695} (\bibinfo{year}{2001}),
  \eprint{astro-ph/0003218}.

\bibitem[{\citenamefont{Medvedev and Loeb}(2013)}]{Medvedev:2012qf}
\bibinfo{author}{\bibfnamefont{M.~V.} \bibnamefont{Medvedev}} \bibnamefont{and}
  \bibinfo{author}{\bibfnamefont{A.}~\bibnamefont{Loeb}},
  \bibinfo{journal}{Mon. Not. Roy. Astron. Soc.}
  \textbf{\bibinfo{volume}{431}}, \bibinfo{pages}{2737} (\bibinfo{year}{2013}),
  \eprint{1212.0333}.

\bibitem[{\citenamefont{Paschalidis et~al.}(2013)\citenamefont{Paschalidis,
  Etienne, and Shapiro}}]{Paschalidis:2013jsa}
\bibinfo{author}{\bibfnamefont{V.}~\bibnamefont{Paschalidis}},
  \bibinfo{author}{\bibfnamefont{Z.~B.} \bibnamefont{Etienne}},
  \bibnamefont{and} \bibinfo{author}{\bibfnamefont{S.~L.}
  \bibnamefont{Shapiro}}, \bibinfo{journal}{Phys. Rev. D}
  \textbf{\bibinfo{volume}{88}}, \bibinfo{pages}{021504}
  (\bibinfo{year}{2013}), \eprint{1304.1805}.

\bibitem[{\citenamefont{Giacomazzo et~al.}(2011)\citenamefont{Giacomazzo,
  Rezzolla, and Baiotti}}]{Giacomazzo:2011ilm}
\bibinfo{author}{\bibfnamefont{B.}~\bibnamefont{Giacomazzo}},
  \bibinfo{author}{\bibfnamefont{L.}~\bibnamefont{Rezzolla}}, \bibnamefont{and}
  \bibinfo{author}{\bibfnamefont{L.}~\bibnamefont{Baiotti}}, in
  \emph{\bibinfo{booktitle}{{46th Rencontres de Moriond on Gravitational Waves
  and Experimental Gravity}}} (\bibinfo{publisher}{Moriond},
  \bibinfo{address}{Paris, France}, \bibinfo{year}{2011}), pp.
  \bibinfo{pages}{69--76}.

\bibitem[{\citenamefont{Antoniadis et~al.}(2013)}]{Antoniadis:2013pzd}
\bibinfo{author}{\bibfnamefont{J.}~\bibnamefont{Antoniadis}}
  \bibnamefont{et~al.}, \bibinfo{journal}{Science}
  \textbf{\bibinfo{volume}{340}}, \bibinfo{pages}{6131} (\bibinfo{year}{2013}),
  \eprint{1304.6875}.

\bibitem[{\citenamefont{Sarin and Lasky}(2021)}]{Sarin:2020gxb}
\bibinfo{author}{\bibfnamefont{N.}~\bibnamefont{Sarin}} \bibnamefont{and}
  \bibinfo{author}{\bibfnamefont{P.~D.} \bibnamefont{Lasky}},
  \bibinfo{journal}{Gen. Rel. Grav.} \textbf{\bibinfo{volume}{53}},
  \bibinfo{pages}{59} (\bibinfo{year}{2021}), \eprint{2012.08172}.

\bibitem[{\citenamefont{Palenzuela
  et~al.}(2013{\natexlab{a}})\citenamefont{Palenzuela, Lehner, Ponce, Liebling,
  Anderson, Neilsen, and Motl}}]{Palenzuela:2013hu}
\bibinfo{author}{\bibfnamefont{C.}~\bibnamefont{Palenzuela}},
  \bibinfo{author}{\bibfnamefont{L.}~\bibnamefont{Lehner}},
  \bibinfo{author}{\bibfnamefont{M.}~\bibnamefont{Ponce}},
  \bibinfo{author}{\bibfnamefont{S.~L.} \bibnamefont{Liebling}},
  \bibinfo{author}{\bibfnamefont{M.}~\bibnamefont{Anderson}},
  \bibinfo{author}{\bibfnamefont{D.}~\bibnamefont{Neilsen}}, \bibnamefont{and}
  \bibinfo{author}{\bibfnamefont{P.}~\bibnamefont{Motl}},
  \bibinfo{journal}{Phys. Rev. Lett.} \textbf{\bibinfo{volume}{111}},
  \bibinfo{pages}{061105} (\bibinfo{year}{2013}{\natexlab{a}}),
  \eprint{1301.7074}.

\bibitem[{\citenamefont{Palenzuela
  et~al.}(2013{\natexlab{b}})\citenamefont{Palenzuela, Lehner, Liebling, Ponce,
  Anderson, Neilsen, and Motl}}]{Palenzuela:2013kra}
\bibinfo{author}{\bibfnamefont{C.}~\bibnamefont{Palenzuela}},
  \bibinfo{author}{\bibfnamefont{L.}~\bibnamefont{Lehner}},
  \bibinfo{author}{\bibfnamefont{S.~L.} \bibnamefont{Liebling}},
  \bibinfo{author}{\bibfnamefont{M.}~\bibnamefont{Ponce}},
  \bibinfo{author}{\bibfnamefont{M.}~\bibnamefont{Anderson}},
  \bibinfo{author}{\bibfnamefont{D.}~\bibnamefont{Neilsen}}, \bibnamefont{and}
  \bibinfo{author}{\bibfnamefont{P.}~\bibnamefont{Motl}},
  \bibinfo{journal}{Phys. Rev. D} \textbf{\bibinfo{volume}{88}},
  \bibinfo{pages}{043011} (\bibinfo{year}{2013}{\natexlab{b}}),
  \eprint{1307.7372}.

\bibitem[{\citenamefont{Abbott et~al.}(2018)}]{KAGRA:2013rdx}
\bibinfo{author}{\bibfnamefont{B.~P.} \bibnamefont{Abbott}}
  \bibnamefont{et~al.} (\bibinfo{collaboration}{KAGRA, LIGO Scientific, Virgo,
  VIRGO}), \bibinfo{journal}{Living Rev. Rel.} \textbf{\bibinfo{volume}{21}},
  \bibinfo{pages}{3} (\bibinfo{year}{2018}), \eprint{1304.0670}.

\bibitem[{\citenamefont{The LIGO Scientific~Collaboration}(2020)}]{ligo20}
\bibinfo{author}{\bibfnamefont{t.~C.} \bibnamefont{The LIGO
  Scientific~Collaboration}, \bibfnamefont{the Virgo~Collaboration}},
  \bibinfo{journal}{Phys. Rev. D. 78, 02412}  (\bibinfo{year}{2020}).

\bibitem[{\citenamefont{Blanchet et~al.}(1995)\citenamefont{Blanchet, Damour,
  Iyer, Will, and Wiseman}}]{Blanchet:1995}
\bibinfo{author}{\bibfnamefont{L.}~\bibnamefont{Blanchet}},
  \bibinfo{author}{\bibfnamefont{T.}~\bibnamefont{Damour}},
  \bibinfo{author}{\bibfnamefont{B.~R.} \bibnamefont{Iyer}},
  \bibinfo{author}{\bibfnamefont{C.~M.} \bibnamefont{Will}}, \bibnamefont{and}
  \bibinfo{author}{\bibfnamefont{A.~G.} \bibnamefont{Wiseman}},
  \bibinfo{journal}{Phys. Rev. Lett.} \textbf{\bibinfo{volume}{74}},
  \bibinfo{pages}{3515} (\bibinfo{year}{1995}),
  \urlprefix\url{https://link.aps.org/doi/10.1103/PhysRevLett.74.3515}.

\bibitem[{\citenamefont{Blanchet}(2007)}]{Blanchet:2006xj}
\bibinfo{author}{\bibfnamefont{L.}~\bibnamefont{Blanchet}},
  \bibinfo{journal}{Comptes Rendus Physique} \textbf{\bibinfo{volume}{8}},
  \bibinfo{pages}{57} (\bibinfo{year}{2007}), \eprint{gr-qc/0611142}.

\bibitem[{\citenamefont{Gergely et~al.}(1998)\citenamefont{Gergely, Perjes, and
  Vasuth}}]{Gergely:1998vp}
\bibinfo{author}{\bibfnamefont{L.~A.} \bibnamefont{Gergely}},
  \bibinfo{author}{\bibfnamefont{Z.}~\bibnamefont{Perjes}}, \bibnamefont{and}
  \bibinfo{author}{\bibfnamefont{M.}~\bibnamefont{Vasuth}}, in
  \emph{\bibinfo{booktitle}{{Spanish Relativity Meeting (ERE 98)}}}
  (\bibinfo{year}{1998}), pp. \bibinfo{pages}{259--262},
  \eprint{gr-qc/9811055}.

\bibitem[{\citenamefont{Gergely}(2000)}]{Gergely:1999pd}
\bibinfo{author}{\bibfnamefont{L.~A.} \bibnamefont{Gergely}},
  \bibinfo{journal}{Phys. Rev. D} \textbf{\bibinfo{volume}{61}},
  \bibinfo{pages}{024035} (\bibinfo{year}{2000}), \eprint{gr-qc/9911082}.

\bibitem[{\citenamefont{Gergely and Keresztes}(2003)}]{Gergely:2002fd}
\bibinfo{author}{\bibfnamefont{L.~A.} \bibnamefont{Gergely}} \bibnamefont{and}
  \bibinfo{author}{\bibfnamefont{Z.}~\bibnamefont{Keresztes}},
  \bibinfo{journal}{Phys. Rev. D} \textbf{\bibinfo{volume}{67}},
  \bibinfo{pages}{024020} (\bibinfo{year}{2003}), \eprint{gr-qc/0211027}.

\bibitem[{\citenamefont{Vasuth et~al.}(2003)\citenamefont{Vasuth, Keresztes,
  Mihaly, and Gergely}}]{Vasuth:2003yr}
\bibinfo{author}{\bibfnamefont{M.}~\bibnamefont{Vasuth}},
  \bibinfo{author}{\bibfnamefont{Z.}~\bibnamefont{Keresztes}},
  \bibinfo{author}{\bibfnamefont{A.}~\bibnamefont{Mihaly}}, \bibnamefont{and}
  \bibinfo{author}{\bibfnamefont{L.~A.} \bibnamefont{Gergely}},
  \bibinfo{journal}{Phys. Rev. D} \textbf{\bibinfo{volume}{68}},
  \bibinfo{pages}{124006} (\bibinfo{year}{2003}), \eprint{gr-qc/0308051}.

\bibitem[{\citenamefont{Dietrich et~al.}(2019)}]{Dietrich:2018uni}
\bibinfo{author}{\bibfnamefont{T.}~\bibnamefont{Dietrich}}
  \bibnamefont{et~al.}, \bibinfo{journal}{Phys. Rev. D}
  \textbf{\bibinfo{volume}{99}}, \bibinfo{pages}{024029}
  (\bibinfo{year}{2019}), \eprint{1804.02235}.

\bibitem[{\citenamefont{Bernuzzi et~al.}(2015)\citenamefont{Bernuzzi, Nagar,
  Dietrich, and Damour}}]{Bernuzzi:2014owa}
\bibinfo{author}{\bibfnamefont{S.}~\bibnamefont{Bernuzzi}},
  \bibinfo{author}{\bibfnamefont{A.}~\bibnamefont{Nagar}},
  \bibinfo{author}{\bibfnamefont{T.}~\bibnamefont{Dietrich}}, \bibnamefont{and}
  \bibinfo{author}{\bibfnamefont{T.}~\bibnamefont{Damour}},
  \bibinfo{journal}{Phys. Rev. Lett.} \textbf{\bibinfo{volume}{114}},
  \bibinfo{pages}{161103} (\bibinfo{year}{2015}), \eprint{1412.4553}.

\bibitem[{\citenamefont{Maggiore}(2007)}]{Maggiore:2007ulw}
\bibinfo{author}{\bibfnamefont{M.}~\bibnamefont{Maggiore}},
  \emph{\bibinfo{title}{{Gravitational Waves. Vol. 1: Theory and
  Experiments}}}, Oxford Master Series in Physics (\bibinfo{publisher}{Oxford
  University Press}, \bibinfo{year}{2007}), ISBN
  \bibinfo{isbn}{978-0-19-857074-5, 978-0-19-852074-0}.

\bibitem[{\citenamefont{Anderson et~al.}(2008)\citenamefont{Anderson,
  Hirschmann, Lehner, Liebling, Motl, Neilsen, Palenzuela, and
  Tohline}}]{Anderson:2008zp}
\bibinfo{author}{\bibfnamefont{M.}~\bibnamefont{Anderson}},
  \bibinfo{author}{\bibfnamefont{E.~W.} \bibnamefont{Hirschmann}},
  \bibinfo{author}{\bibfnamefont{L.}~\bibnamefont{Lehner}},
  \bibinfo{author}{\bibfnamefont{S.~L.} \bibnamefont{Liebling}},
  \bibinfo{author}{\bibfnamefont{P.~M.} \bibnamefont{Motl}},
  \bibinfo{author}{\bibfnamefont{D.}~\bibnamefont{Neilsen}},
  \bibinfo{author}{\bibfnamefont{C.}~\bibnamefont{Palenzuela}},
  \bibnamefont{and} \bibinfo{author}{\bibfnamefont{J.~E.}
  \bibnamefont{Tohline}}, \bibinfo{journal}{Phys. Rev. Lett.}
  \textbf{\bibinfo{volume}{100}}, \bibinfo{pages}{191101}
  (\bibinfo{year}{2008}), \eprint{0801.4387}.

\bibitem[{\citenamefont{Gruzinov}(2007)}]{Gruzinov:2007qa}
\bibinfo{author}{\bibfnamefont{A.}~\bibnamefont{Gruzinov}},
  \bibinfo{journal}{Astrophys. J. Lett.} \textbf{\bibinfo{volume}{667}},
  \bibinfo{pages}{L69} (\bibinfo{year}{2007}), \eprint{astro-ph/0702243}.

\bibitem[{\citenamefont{Cerutti and Beloborodov}(2017)}]{Cerutti:2016ttn}
\bibinfo{author}{\bibfnamefont{B.}~\bibnamefont{Cerutti}} \bibnamefont{and}
  \bibinfo{author}{\bibfnamefont{A.}~\bibnamefont{Beloborodov}},
  \bibinfo{journal}{Space Sci. Rev.} \textbf{\bibinfo{volume}{207}},
  \bibinfo{pages}{111} (\bibinfo{year}{2017}), \eprint{1611.04331}.

\bibitem[{\citenamefont{Lattimer}(2012)}]{Lattimer:2012nd}
\bibinfo{author}{\bibfnamefont{J.~M.} \bibnamefont{Lattimer}},
  \bibinfo{journal}{Ann. Rev. Nucl. Part. Sci.} \textbf{\bibinfo{volume}{62}},
  \bibinfo{pages}{485} (\bibinfo{year}{2012}), \eprint{1305.3510}.

\bibitem[{\citenamefont{Demorest et~al.}(2010)\citenamefont{Demorest, Pennucci,
  Ransom, Roberts, and Hessels}}]{Demorest:2010bx}
\bibinfo{author}{\bibfnamefont{P.}~\bibnamefont{Demorest}},
  \bibinfo{author}{\bibfnamefont{T.}~\bibnamefont{Pennucci}},
  \bibinfo{author}{\bibfnamefont{S.}~\bibnamefont{Ransom}},
  \bibinfo{author}{\bibfnamefont{M.}~\bibnamefont{Roberts}}, \bibnamefont{and}
  \bibinfo{author}{\bibfnamefont{J.}~\bibnamefont{Hessels}},
  \bibinfo{journal}{Nature} \textbf{\bibinfo{volume}{467}},
  \bibinfo{pages}{1081} (\bibinfo{year}{2010}), \eprint{1010.5788}.

\bibitem[{\citenamefont{{Cheng} and {Dai}}(1997)}]{1997ApJ...476L..39C}
\bibinfo{author}{\bibfnamefont{K.~S.} \bibnamefont{{Cheng}}} \bibnamefont{and}
  \bibinfo{author}{\bibfnamefont{Z.~G.} \bibnamefont{{Dai}}},
  \bibinfo{journal}{APJL} \textbf{\bibinfo{volume}{476}}, \bibinfo{pages}{L39}
  (\bibinfo{year}{1997}).

\bibitem[{\citenamefont{Jackson}(1998)}]{Jackson:1998nia}
\bibinfo{author}{\bibfnamefont{J.~D.} \bibnamefont{Jackson}},
  \emph{\bibinfo{title}{{Classical Electrodynamics}}}
  (\bibinfo{publisher}{Wiley}, \bibinfo{year}{1998}), ISBN
  \bibinfo{isbn}{978-0-471-30932-1}.

\bibitem[{\citenamefont{Liu et~al.}(2008)\citenamefont{Liu, Shapiro, Etienne,
  and Taniguchi}}]{Liu:2008xy}
\bibinfo{author}{\bibfnamefont{Y.~T.} \bibnamefont{Liu}},
  \bibinfo{author}{\bibfnamefont{S.~L.} \bibnamefont{Shapiro}},
  \bibinfo{author}{\bibfnamefont{Z.~B.} \bibnamefont{Etienne}},
  \bibnamefont{and}
  \bibinfo{author}{\bibfnamefont{K.}~\bibnamefont{Taniguchi}},
  \bibinfo{journal}{Phys. Rev. D} \textbf{\bibinfo{volume}{78}},
  \bibinfo{pages}{024012} (\bibinfo{year}{2008}), \eprint{0803.4193}.

\bibitem[{\citenamefont{Mikoczi et~al.}(2005)\citenamefont{Mikoczi, Vasuth, and
  Gergely}}]{Mikoczi:2005dn}
\bibinfo{author}{\bibfnamefont{B.}~\bibnamefont{Mikoczi}},
  \bibinfo{author}{\bibfnamefont{M.}~\bibnamefont{Vasuth}}, \bibnamefont{and}
  \bibinfo{author}{\bibfnamefont{L.~A.} \bibnamefont{Gergely}},
  \bibinfo{journal}{Phys. Rev. D} \textbf{\bibinfo{volume}{71}},
  \bibinfo{pages}{124043} (\bibinfo{year}{2005}), \eprint{astro-ph/0504538}.

\bibitem[{\citenamefont{Lincoln and Will}(1990)}]{Lincoln:1990ji}
\bibinfo{author}{\bibfnamefont{C.~W.} \bibnamefont{Lincoln}} \bibnamefont{and}
  \bibinfo{author}{\bibfnamefont{C.~M.} \bibnamefont{Will}},
  \bibinfo{journal}{Phys. Rev. D} \textbf{\bibinfo{volume}{42}},
  \bibinfo{pages}{1123} (\bibinfo{year}{1990}).

\bibitem[{\citenamefont{Blanchet et~al.}(1996)\citenamefont{Blanchet, Iyer,
  Will, and Wiseman}}]{Blanchet:1996pi}
\bibinfo{author}{\bibfnamefont{L.}~\bibnamefont{Blanchet}},
  \bibinfo{author}{\bibfnamefont{B.~R.} \bibnamefont{Iyer}},
  \bibinfo{author}{\bibfnamefont{C.~M.} \bibnamefont{Will}}, \bibnamefont{and}
  \bibinfo{author}{\bibfnamefont{A.~G.} \bibnamefont{Wiseman}},
  \bibinfo{journal}{Class. Quant. Grav.} \textbf{\bibinfo{volume}{13}},
  \bibinfo{pages}{575} (\bibinfo{year}{1996}), \eprint{gr-qc/9602024}.

\bibitem[{\citenamefont{Blanchet}(2014)}]{Blanchet:2013haa}
\bibinfo{author}{\bibfnamefont{L.}~\bibnamefont{Blanchet}},
  \bibinfo{journal}{Living Rev. Rel.} \textbf{\bibinfo{volume}{17}},
  \bibinfo{pages}{2} (\bibinfo{year}{2014}), \eprint{1310.1528}.

\bibitem[{\citenamefont{Shapiro and Teukolsky}(1983)}]{Shapiro:1983du}
\bibinfo{author}{\bibfnamefont{S.~L.} \bibnamefont{Shapiro}} \bibnamefont{and}
  \bibinfo{author}{\bibfnamefont{S.~A.} \bibnamefont{Teukolsky}},
  \emph{\bibinfo{title}{{Black holes, white dwarfs, and neutron stars: The
  physics of compact objects}}} (\bibinfo{publisher}{WILEY-VCH},
  \bibinfo{year}{1983}), ISBN \bibinfo{isbn}{978-0-471-87316-7}.

\bibitem[{\citenamefont{Abbott et~al.}(2021)}]{LIGOScientific:2021djp}
\bibinfo{author}{\bibfnamefont{R.}~\bibnamefont{Abbott}} \bibnamefont{et~al.}
  (\bibinfo{collaboration}{LIGO Scientific, VIRGO, KAGRA})
  (\bibinfo{year}{2021}), \eprint{2111.03606}.

\bibitem[{\citenamefont{Apostolatos}(1994)}]{Apostolatos:1994ha}
\bibinfo{author}{\bibfnamefont{T.~A.} \bibnamefont{Apostolatos}}, in
  \emph{\bibinfo{booktitle}{{7th Marcel Grossmann Meeting on General Relativity
  (MG 7)}}} (\bibinfo{year}{1994}), pp. \bibinfo{pages}{1075--1077}.

\bibitem[{\citenamefont{Capozziello et~al.}(2011)\citenamefont{Capozziello,
  Laurentis, Martino, Formisano, and Vernieri}}]{Capozziello:2011fs}
\bibinfo{author}{\bibfnamefont{S.}~\bibnamefont{Capozziello}},
  \bibinfo{author}{\bibfnamefont{M.~D.} \bibnamefont{Laurentis}},
  \bibinfo{author}{\bibfnamefont{I.~D.} \bibnamefont{Martino}},
  \bibinfo{author}{\bibfnamefont{M.}~\bibnamefont{Formisano}},
  \bibnamefont{and} \bibinfo{author}{\bibfnamefont{D.}~\bibnamefont{Vernieri}},
  \bibinfo{journal}{Astrophys. Space Sci.} \textbf{\bibinfo{volume}{333}},
  \bibinfo{pages}{29} (\bibinfo{year}{2011}), \eprint{1101.5306}.

\bibitem[{\citenamefont{Hannam et~al.}(2014)\citenamefont{Hannam, Schmidt,
  Boh\'e, Haegel, Husa, Ohme, Pratten, and P\"urrer}}]{Hannam:2013oca}
\bibinfo{author}{\bibfnamefont{M.}~\bibnamefont{Hannam}},
  \bibinfo{author}{\bibfnamefont{P.}~\bibnamefont{Schmidt}},
  \bibinfo{author}{\bibfnamefont{A.}~\bibnamefont{Boh\'e}},
  \bibinfo{author}{\bibfnamefont{L.}~\bibnamefont{Haegel}},
  \bibinfo{author}{\bibfnamefont{S.}~\bibnamefont{Husa}},
  \bibinfo{author}{\bibfnamefont{F.}~\bibnamefont{Ohme}},
  \bibinfo{author}{\bibfnamefont{G.}~\bibnamefont{Pratten}}, \bibnamefont{and}
  \bibinfo{author}{\bibfnamefont{M.}~\bibnamefont{P\"urrer}},
  \bibinfo{journal}{Phys. Rev. Lett.} \textbf{\bibinfo{volume}{113}},
  \bibinfo{pages}{151101} (\bibinfo{year}{2014}), \eprint{1308.3271}.

\bibitem[{\citenamefont{Zhang and Meszaros}(2001)}]{Zhang:2000wx}
\bibinfo{author}{\bibfnamefont{B.}~\bibnamefont{Zhang}} \bibnamefont{and}
  \bibinfo{author}{\bibfnamefont{P.}~\bibnamefont{Meszaros}},
  \bibinfo{journal}{Astrophys. J. Lett.} \textbf{\bibinfo{volume}{552}},
  \bibinfo{pages}{L35} (\bibinfo{year}{2001}), \eprint{astro-ph/0011133}.

\end{thebibliography}

%%%%%%%%%%%%%%%
%%%   END   %%%
%%%%%%%%%%%%%%%

\end{document}